\documentclass[preprint,12pt]{elsarticle}
\usepackage{amssymb, amsmath, mathrsfs}

\usepackage{amsthm}
\theoremstyle{definition}
\newtheorem{definition}{Definition}[section]

\newtheorem{remark}{Remark}[section]
\theoremstyle{plain}
\newtheorem{theorem}{Theorem}[section]

\journal{
Potential Analysis
}

\begin{document}

\begin{frontmatter}

\title{Formulas that represent Cauchy problem solution for momentum and position Schr\"{o}dinger equation}

\author{Ivan\,D.\,Remizov}

\address{National Research University Higher School of Economics
	
	25/12 Bol. Pecherskaya Ulitsa, Room 224, Nizhny Novgorod, 603155, Russia
}

\ead{ivremizov@yandex.ru}

\begin{abstract}
In the paper we derive two formulas representing solutions of Cauchy problem for two Schr\"{o}dinger equations: one-dimensional momentum space equation with polynomial potential, and  multidimensional position space equation with locally square integrable potential. The first equation is a constant coefficients particular case of an evolution equation with derivatives of arbitrary high order and variable coefficients that do not change over time, this general equation is solved in the paper. We construct a family of translation operators in the space of square integrable functions and then use methods of functional analysis based on Chernoff product formula to prove that this family approximates the solution-giving semigroup. This leads us to some formulas that express the solution for Cauchy problem in terms of initial condition and coefficients of the equations studied.

\end{abstract}


\end{frontmatter}

\footnotesize
\tableofcontents           
\normalsize
			
\section{Introduction}
\subsection{Motivation}

One can write the Schr\"odinger equation for a particle with the mass $m$ in the potential $V$ as follows:
\begin{equation}\label{schrmompos}
i\hbar \psi'_t=H\psi,\quad \mathrm{ where  }\quad H(\hat{p},\hat{q})=\frac{1}{2m}\hat{p}^2 + V(\hat{q}).
\end{equation}

Below we use auxiliary variable $x$ to write $\psi(t,q)$ and $\psi(t,p)$ in a unified way as $\psi(t,x)$ and hope that it will not cause misunderstanding. 

\subsubsection{Momentum Schr\"odinger equation}

If one knows the initial state $\psi_0$ for the momentum of the particle, then it is possible to predict the momentum in all the future and the past via the following algorythm.

1. Set $\hat{q}=-i\hbar \frac{\partial}{\partial x}$, i.e. $(\hat{q}f)(x)=-i\hbar f'(x)$.

2. Set $\hat{p}=x$, i.e. $(\hat{p}f)(x)=x f(x)$.

3. Write the Cauchy problem for equation  (\ref{schrmompos}) in the form

\begin{equation}\label{CPmom}
\left\{ \begin{array}{ll}
i\hbar\psi'_t(t,x)=\frac{1}{2m}x^2\psi(t,x) + V(-i\hbar \frac{\partial}{\partial x})\psi(t,x),\\ 
\psi(0,x)=\psi_0(x).
\end{array} \right.
\end{equation}

For example, if $V(x)=x^2 + x^4$ then $V(-i\hbar \frac{\partial}{\partial x})\psi(t,x)=-\hbar^2\psi''_{xx}(t,x)+\hbar^4\psi''''_{xxxx}(t,x)$. If $V$ is not a polynomial then $V(-i\hbar \frac{\partial}{\partial x})$ is a pseudo-differential operator which can be defined via the Fourier transform.

4. Solve this Cauchy problem, i.e. find $\psi(t,x)$ for all $t$ and $x$.

5. Come up with a set $A$ in the momentum space of the particle. Then the probability that the particle in time $t$ has a momentum in the set $A$ is equal to  $\int_A|\psi(t,x)|^2dx$.

In the present paper we solve the Chauchy problem (\ref{CP1}) which covers the case (\ref{CPmom}) for $x\in\mathbb{R}^1$ and a  polynomial potential $V$. In fact, problem (\ref{CP1}) is more general than (\ref{CPmom}). The difference is that the coefficients of the polynomial can be variable, and $\frac{1}{2m}x^2$ can be substituted by any measurable  square-integrable function $a_0(x)$, see theorem \ref{mainth} for the deatils. See  \cite{M1, M2, Buz2017} and references therein for known results related to  Cauchy problems for evolution equations with derivatives of higher orders. See also \cite{Maz, Krav2012, KSS2016} and references therein for equations with polynomial potential.

\subsubsection{Position Schr\"odinger equation}

Similarly, if one knows the initial state $\psi_0$ for the position of the particle, then it is possible to predict the position in all the future and the past via the following algorythm.

1. Set $\hat{p}=i\hbar \frac{\partial}{\partial x}$, i.e. $(\hat{p}f)(x)=i\hbar f'(x)$.

2. Set $\hat{q}=x$, i.e. $(\hat{q}f)(x)=x f(x)$.

3. Write the Cauchy problem for equation  (\ref{schrmompos}) in the form

\begin{equation}\label{CPpos}
\left\{ \begin{array}{ll}
i\hbar\psi'_t(t,x)=-\frac{1}{2m}\hbar^2\psi''_{xx}(t,x) + V(x)\psi(t,x),\\ 
\psi(0,x)=\psi_0(x).
\end{array} \right.
\end{equation}

4. Solve this Cauchy problem, i.e. find $\psi(t,x)$ for all $t$ and $x$.

5. Come up with a set $A$ in the position space of the particle. Then the probability that the particle in time $t$ has a position in the set $A$ is equal to  $\int_A|\psi(t,x)|^2dx$.

In the present paper for $x\in\mathbb{R}^d$, $d=1,2,3,\dots$ we solve the Chauchy problem (\ref{f_Koshi}) which is equivalent to (\ref{CPpos}), see theorem \ref{thRd} for the details. See also Chapter 11 in \cite{JL} and references therein for known results related to such Cauchy problems.

\subsection{Problem setting and approach proposed}

A relatively small number of examples is known where the solution of a differential equation with variable coefficients can be expressed (more or less) explicitly via some formula in terms of these coefficients. In this paper we provide such formulas for the Schr\"odinger equation. Most of the paper is devoted to studying the one-dimensional case, but in the last chapter the multi-dimensional case is considered. Let us first describe the equations and then provide the necessary background.

\subsubsection{One-dimensional case}
 
For fixed $K\in\mathbb{N}$ we study the following Cauchy problem for Schr\"o\-din\-ger equation (which in this case is a partial differential equation of order $2K$)
\begin{equation}\label{CP1}
\left\{ \begin{array}{ll}
i\frac{\partial}{\partial t}\psi(t,x)=\sum\limits_{k=0}^{K}\frac{\partial^k}{\partial x^k}\left(a_k(x) \frac{\partial^k}{\partial x^k} \psi(t,x)\right)\stackrel{\textrm{denote}}{=}\mathcal{H}\psi(t,x);\quad x\in\mathbb{R}^1, t\geq 0,\\ 
\psi(0,x)=\psi_0(x);\quad x\in\mathbb{R}^1,
\end{array} \right.
\end{equation}
where for $k=1,\dots,K$ coefficients $a_k$ are bounded smooth functions $a_k\colon\mathbb{R}\to\mathbb{R}$ with bounded derivatives up to $(2k)$-th order,  while the coefficient $a_0\colon\mathbb{R}\to\mathbb{R}$ is measurable but may be unbounded (see theorem \ref{mainth} for all technical details). We also assume that coefficients $a_k$, $k=0,1,\dots,K$ are chosen in such a way that operator $\mathcal{H}$ is self-adjoint and  defined on some dense linear subspace of $L_2(\mathbb{R})$. The initial condition $\psi_0\colon\mathbb{R}\to\mathbb{C}$ belongs to a complex Lebesgue space $L_2(\mathbb{R})$ which is a Hilbert space over the field $\mathbb{C}$. The fact that $\psi_0\in L_2(\mathbb{R})$ means that $\psi_0\colon \mathbb{R}\to \mathbb{C}$ is a measurable function and  $\int_{-\infty}^{+\infty} |\psi_0(x)|^2dx<\infty$ with respect to the Lebesgue measure on the real line $(-\infty,+\infty)$. As usual, we say that two functions represent the same vector of $L_2(\mathbb{R})$ iff they are equal almost everywhere; one can find the definition of $L_2(\mathbb{R})$ space and corresponding facts of measure\&integral theory in  \cite{KF}. The right-hand side of the first equation in (\ref{CP1}) determines a densely defined self-adjoint operator in $L_2(\mathbb{R})$, which is in line with the physical meaning of the Schr\"odinger equation. Moreover, it is known \cite{Naim} that any self-adjoint differential expression on the real line with real variable coefficients (which together with the domain define the operator) has even order and there exist such functions $a_k$ that the expression could be represented in the form of the right-hand side of the first  equation in (\ref{CP1}). So the case considered appears to be general for equations with real coefficients. However, there are known self-adjoint differential expressions of odd order with coefficients with non-zero imaginary part \cite{Naim}, which we do not discuss in the present paper.

We want to find a solution $\psi$ such that for each $t\geq 0$ we have $\psi(t,\cdot)\in L_2(\mathbb{R})$ and (\ref{CP1}) is satisfied in sence of $L_2(\mathbb{R})$. This solution is known to exist for each $\psi_0\in L_2(\mathbb{R})$ and is provided by the resolving $C_0$-semigropup for the equation considered because the operator on the right-hand side of the equation is self-adjoint \cite{Stone, EN1}. But even being sure of the existence (and in some classes of functions -- of the uniqueness) of the solution, we are still curious to find a formula that expresses the solution of (\ref{CP1}) in terms of coefficients of (\ref{CP1}); this paper provides such formula. We employ general approach proposed in \cite{R1} to find an explicit formula for the resolving $C_0$-semigroup and thus reaching the proposed goal. The result with full details is given in theorem \ref{mainth}.

\subsubsection{Multi-dimensional case}

In the last chapter, for arbitrary fixed $d\in\mathbb{N}$ we obtain the solution of the Cauchy problem for a $d$-dimensional Schr\"odinger equation. In the space  $L_2(\mathbb{R}^d)$ over the field $\mathbb{C}$ we study a problem
\begin{equation}\label{f_Koshi}
\begin{cases}
\psi'_{t}(t,x)=\frac{1}{2}i\left(\sum\limits_{m=1}^d\psi_{x_mx_m}''(t,x)\right)-iV(x)\psi(t,x), & t\in\mathbb{R}^1, x\in\mathbb{R}^d, \\
\psi(0,x)=\psi_{0}(x), & x\in\mathbb{R}^d.
\end{cases}
\end{equation} 
We assume that function   $V\colon\mathbb{R}^d\to \mathbb{R}$ is measurable and has a locally summable second power, $V\in L_2^{loc}(\mathbb{R}^d)$. For example, $V$ can be an arbitrary continuous non-negative function, including cases of quantum harmonic oscillator ($V(x)=\|x\|^2$) and two most known quantum anharmonic oscillators  ($V(x)=\|x\|^4$, $V(x)=\|x\|^2+\|x\|^4$). The result with full details is given in theorem \ref{thRd}.

Now let us provide some background in the field, sketch heuristic arguments to explain the idea of our method without technical formalities, and finally state and prove theorems.

\subsection{$C_0$-semigroups and linear evolution equations}\label{semgreveq}  

Let us provide a very short  introduction to $C_0$-semigroup theory and show its connection to linear evolution equations in general and with the Cauchy problem for the Scr\"odinger equation in particular. One can find proofs and other details in monograph \cite{EN1}.

\begin{definition}\label{semigrdef}
Let $\mathcal{F}$ be a Banach space over the field $\mathbb{C}$. Let $\mathscr{L}(\mathcal{F})$ be a set of all bounded linear operators in $\mathcal{F}$. Suppose we have a mapping $V\colon [0,+\infty)\to \mathscr{L}(\mathcal{F}),$ i.e. $V(t)$ is a bounded linear operator $V(t)\colon \mathcal{F}\to \mathcal{F}$ for each $t\geq 0.$ The mapping $V$ is called a \textit{$C_0$-semigroup}, or \textit{a strongly continuous one-parameter semigroup} if it satisfies the following conditions: 
	
1) $V(0)$ is the identity operator $I$, i.e. $\forall \varphi\in \mathcal{F}: V(0)\varphi=\varphi;$ 
	
2) $V$ maps the addition of numbers in $[0,+\infty)$ into a composition of operators in $\mathscr{L}(\mathcal{F})$, i.e. $\forall t\geq 0,\forall s\geq 0: V(t+s)=V(t)\circ V(s),$ where $(A\circ B)(\varphi)=A(B(\varphi))=AB\varphi$;
	
3) $V$ is continuous with respect to the strong operator topology in $\mathscr{L}(\mathcal{F})$, i.e. $\forall \varphi\in \mathcal{F}$ function $t\longmapsto V(t)\varphi$ is continuous as a mapping $[0,+\infty)\to \mathcal{F}.$
	
The definition of a \textit{$C_0$-group} is obtained by substitution of $[0,+\infty)$ with $\mathbb{R}$ in the paragraph above.
\end{definition}

It is known \cite{EN1} that if $(V(t))_{t\geq 0}$ is a $C_0$-semigroup in Banach space $\mathcal{F}$, then the set $$\left\{\varphi\in \mathcal{F}: \exists \lim_{t\to +0}\frac{V(t)\varphi-\varphi}{t}\right\}\stackrel{denote}{=}Dom(L)$$ is dense in $\mathcal{F}$. The operator $L$ defined on the domain $Dom(L)$ by the equality $$L\varphi=\lim_{t\to +0}\frac{V(t)\varphi-\varphi}{t}$$ is called \textit{an infinitesimal generator} (or just \textit{generator} for short) of the $C_0$-semigroup $(V(t))_{t\geq 0}$. The generator is a closed linear operator that defines the $C_0$-semigroup uniquely, which is denoted as $V(t)=e^{tL}$. If $L$ is a bounded operator and $Dom(L)=\mathcal{F}$, then $e^{tL}$ is indeed the exponent defined by the power series $e^{tL}=\sum_{k=0}^\infty\frac{t^kL^k}{k!}$ converging with respect to the norm topology in $\mathscr{L}(\mathcal{F})$. In most interesting cases the generator is an unbounded differential operator such as Laplacian $\Delta$.

One of the reasons for the study of $C_0$-semigroups is their connection with differential equations. If $Q$ is a set, then the function $u\colon [0,+\infty)\times Q\to \mathbb{C}$, $u\colon (t,x)\longmapsto u(t,x)$ of two variables $(t,x)$ can be considered as a function $u\colon t\longmapsto [x\longmapsto u(t,x)]$ of one variable $t$
with values in the space of functions of variable $x$. If $u(t,\cdot)\in\mathcal{F}$ then one can define $Lu(t,x)=(Lu(t,\cdot))(x).$ If there exists a $C_0$-semigroup $(e^{tL})_{t\geq 0}$ then the Cauchy problem 
\begin{equation}\label{CPee}
\left\{ \begin{array}{ll}
u'_t(t,x)=Lu(t,x) \ \mathrm{ for }\ t>0, x\in Q\\
u(0,x)=u_0(x)\ \mathrm{ for } \ x\in Q
\end{array} \right.
\end{equation}
has a unique (in sense of $\mathcal{F}$, where $u(t,\cdot)\in\mathcal{F}$ for every $t\geq 0$) solution $u(t,x)=(e^{tL}u_0)(x)$ which depends on $u_0$ continuously. See also different meanings of the solution \cite{EN1} (including the mild solution which solves the corresponding integral equation). Note that if there exists a strongly continuous group $(e^{tL})_{t\in\mathbb{R}}$ then in the Cauchy problem the equation $u'_t(t,x)=Lu(t,x)$ can be considered not only for $t>0$, but for $t\in\mathbb{R}$, and the solution is provided by the same formula $u(t,x)=(e^{tL}u_0)(x)$.

The equation $u'_t(t,x)=Lu(t,x)$ is called \textit{a linear evolution equation} reflecting the fact that  the operator $L$ is linear. Note that Cauchy problems (\ref{CP1}), (\ref{f_Koshi}) and (\ref{CPgen}) belong to class (\ref{CPee}), i.e. Schr\"odinger equation is a linear evolution equation. This allows us to use the technique of $C_0$-semigrops to reach the main goal of the paper.

The following theorem together with the above theory implies the existence and uniqueness of the solution for the Cauchy problem for the Schr\"{o}din\-ger equation (\ref{CPgen}).

\begin{theorem}\label{Stth} (\textsc{M.\,H.~Stone, 1932}; cf. original paper \cite{Stone} and  theorem 3.24 in \cite{EN1}.) There is a one-to-one correspondence between the linear self-adjoint operators $A$ in Hilbert space $\mathcal{F}$ and the unitary strongly continuous groups $(U(t))_{t\in \mathbb{R}}$ of linear bounded operators in $\mathcal{F}$. 
	
This correspondence is the following: $iA$ is the generator of $(U(t))_{t\in \mathbb{R}}$, which is denoted as $U(t)=e^{itA}.$ 
\end{theorem}

\subsection{Chernoff theorem and Chernoff functions} 

\begin{definition}\label{CTdef} (First introduced in \cite{R1})
Let us say that $G$ is \textit{Chernoff-tangent} to $L$ iff the following conditions of Chernoff tangency (CT) hold: 

(CT0). Let $\mathcal{F}$ be a Banach space, and $\mathscr{L}(\mathcal{F})$ be a space of all linear bounded operators in $\mathcal{F}$. Suppose that we have an operator-valued function $G\colon [0, +\infty) \to \mathscr{L}(\mathcal{F})$, or, using other words, we have a family $(G(t))_{t\geq 0}$ of linear bounded operators in $\mathcal{F}$. Closed linear operator $L\colon Dom(L) \to \mathcal{F}$ is defined on the linear subspace $Dom(L)\subset\mathcal{F}$ which is dense in  $\mathcal{F}$.

(CT1). Function $G$ is strongly continuous, i.e. continuous in the strong topology in  $\mathscr{L}(\mathcal{F})$; in other words, the mapping $t \longmapsto G(t)f\in\mathcal{F}$ is continuous on $[0, +\infty)$ for each $f \in \mathcal{F}$;

(CT2). $G(0) = I,$ i.~e. $G(0)f=f$ for each $f \in \mathcal{F}$;

(CT3). There exists such a linear subspace $\mathcal{D} \subset \mathcal{F}$ that it is dense in  $\mathcal{F}$ and for each $f \in \mathcal{D}$ there exists a limit $\lim_{t \to 0}(G(t)f-f)/t$; let us denote the value of this limit as   $G'(0)f$;

(CT4). Closure of the operator $(G'(0), \mathcal{D})$ exists and is equal to $(L, Dom(L))$.
\end{definition}

\begin{theorem}\label{FormulaChernova}(\textsc{P.\,R.~Chernoff, 1968}; cf. original paper  \cite{Chernoff}, theorem 5.2 in  \cite{EN1} and theorem 10.7.21 in \cite{BS}.) In the notation of the above definition suppose that $L$ and $G$ satisfy:
	
(E). There exists a $C_0$-semigroup $(e^{tL})_{t\geq 0}$ and its generator is $(L,Dom(L))$.
	
(CT). The function $G$ is Chernoff-tangent to operator $(L,Dom(L))$.

(N). There exists $\omega\in\mathbb{R}$ such that $\|G(t)\|\leq e^{\omega t}$ for all $t\geq 0$.

Then for each $f\in \mathcal{F}$ and each $T>0$ we have 
\begin{equation}\label{Chest}
\lim_{n\to\infty}\sup_{t\in[0,T]}\left\|G(t/n)^nf - e^{tL}f\right\|=0,
\end{equation}
where $G(t/n)^n$ is a composition of $n$ copies of linear bounded operator $G(t/n)$.
\end{theorem}

\begin{remark}
If $G$ is Chernoff-tangent to $L$, then the expression $G(t/n)^nf$ is called \textit{a Chernoff approximation expression} for $e^{tL}f$, and $G(t/n)^nu_0$ is called \textit{a Chernoff formal solution} for Cauchy problem $[u'(t)=Lu(t); u(0)=u_0]$. If, moreveover, (\ref{Chest}) holds, then $G$ is called \textit{a Chernoff function} for operator $L$, and $G$ is called \textit{Chernoff-equivalent} to $C_0$-semigroup $(e^{tL})_{t\geq 0}$; in this case $u(t)=U(t)u_0=\lim_{n\to\infty}G(t/n)^nu_0=e^{tL}u_0$ can be shown to be a solution of this Cauchy problem.
\end{remark}

\begin{remark}
The Chernoff theorem (and definitions derived from it) admit two equivalent wordings: with unbounded time and with arbitrary small time. The first is provided above. The second arises when Chernoff function in (CT) is defined not for all $t\geq 0$, but only for $t\in[0,\delta)$ for fixed small $\delta>0$. The condition $(N)$ is substituted by the following condition $(N')$: 

\textit{$(N')$ There exists $\alpha>0$ such that $\|G(t)\|\leq 1+\alpha t$ for all $t\in[0,\delta)$.}

This wording is motivated by the fact that the value of $t/n$ in Chernoff approximation expression $G(t/n)^nf$  becomes arbitrary small as $n\to\infty$ while $t\in[0,T]$. This is also in line with the condition (CT3) which itself uses $G(t)$ defined only for small values of $t>0$.
\end{remark}

\begin{remark}
One may ask why finding Chernoff function for operator $L$ is simpler than finding $e^{tL}$ using some other method? Why one should use Chernoff's theorem? The first reason is that there are no standard methods for most important operators $L$ with variable coefficients, so usually we can only refer to solving the Cauchy problem (\ref{CPee}) for each $u_0\in\mathcal{F}$. The second reason is that Chernoff function $G$ may not have semigroup composition property ($G(t_1+t_2)\neq G(t_1)G(t_2)$), which gives us some freedom in writing the formula for $G$, allowing for a shorter and simplier formulation.  
\end{remark}

\begin{remark}
The definition of Chernoff equivalence goes back to 2002's definiton by O.G.~Smo\-lya\-nov \cite{STT}, who since 2000's papers \cite{SWWdan, SWWcan, STmzm} systematically applies Chernoff's theorem to solving Cauchy problem for linear evolution equations, see overviews \cite{SmHist, SmSchrHist} and shorter in the section that follows below. 
\end{remark}

\begin{remark}
An important question is how fast the error decreases in the approximation expression provided by the Chernoff theorem as $n$ tends to infinity, and the same question for the Trotter product formula $e^{A+B}=\lim_{n\to\infty}(e^{A/n}e^{B/n})^n$. The research here is far from the endinig, several recent papers are \cite{Zag-1, Zag-2, Zag-3}.
\end{remark}

\subsection{Schr\"odinger Equation and Quantum Mechanics} 

Schr\"odinger equation is one of the main equations of Quntum Mechanics  \cite{Town, BerSh, Mull}. When the coefficients of the Schr\"odinger equation do not depend on time, the equation describes the evolution of a closed quantum system, i.e. how the system changes over time under the condition of the system being isolated (not interacting with any external particles or fields). If the quantum system is obtained via quantization of some classical system with configuration space $Q$, then pure state $\psi$ of the quantum system is a vector of unit length ($\|\psi\|=1$) that belongs to complex Hilbert space $L_2(Q)$. As vectors of $L_2(Q)$ are functions $\psi\colon Q\to\mathbb{C}$, pure state $\psi$ is also called a wave function. This terminology has physical meaning, which we do not discuss instead directing the reader to \cite{Town, BerSh, Mull}.

In the process of evolution pure states go to pure states. As time goes from $0$ to $t$, evolution of the system from the initial pure state $\psi(0)=\psi_0\in L_2(Q)$ to pure state $\psi(t)\in L_2(Q)$ can be described as applying linear bounded unitary operator $U(t)$ to $\psi_0$, i.~e. $\psi(t)=U(t)\psi_0$. As the operator $U(t)$ is unitary, we have $\|\psi(t)\|=\left\|U(t)\psi_0\right\|=\|\psi_0\|=1$ which is in line with proceeding from one pure state to another pure state. The evolution operator $U(t)$ is connected to the Hamiltonian $\mathcal{H}$ of the system by the relation  $U(t)=e^{-it\mathcal{H}}$. The Hamiltonian describes pure state $\psi(t)$ via Cauchy problem for the Scr\"odinger equation  
\begin{equation}\label{CPgen}
\left\{ \begin{array}{ll}
i\psi'_t(t)=\mathcal{H}\psi(t),\\ 
\psi(0)=\psi_0.
\end{array} \right.
\end{equation}

In general case (i.e. for arbitrary quantum system) the Hamiltonian $\mathcal{H}$ is a self-adjoint operator in $L_2(Q)$ with dense domain $Dom(\mathcal{H})\subset L_2(Q)$; theory of such operators can be found in \cite{EN1, BS, Yos, RS}. The condition of being self-adjoint is very important: it guarantees (thanks to the Stone theorem \cite{Stone, EN1}, see theorem \ref{Stth} above) that for each $t\in\mathbb{R}$ the operator $e^{-it\mathcal{H}}$ exists and can be shown to be unitary. Moreover, the family $\left(e^{-it\mathcal{H}}\right)_{t\in\mathbb{R}}$ can be shown to be a one-parameter strongly continuous group (or a $C_0$-group for short) of unitary linear bounded operators with infinitesimal generator $-i\mathcal{H}$ \cite{EN1}. The Cauchy problem (\ref{CPgen}) then has a unique solution provided by the formula $\psi(t)=e^{-it\mathcal{H}}\psi_0$. 

Summing up what has been said, if we want to determine the evolution of a quantum system we need to determine either the Hamiltonian $\mathcal{H}$ and find $\psi(t)$ from the Cauchy problem (\ref{CPgen}) or the evolution operator $U(t)$ for each $t\in\mathbb{R}$ and find $\psi(t)$ via formula $\psi(t)=U(t)\psi_0$. Both variants bring us to the same result $\psi(t)=U(t)\psi_0=e^{-it\mathcal{H}}\psi_0$. Usually, the Hamiltonian is known and the evolution operator is not. Unfortunatelly, even if we know $\mathcal{H}$ the formula $U(t)=e^{-it\mathcal{H}}$ is not usable for direct calculation of  $U(t)$ when the operator $\mathcal{H}$ is not bounded, which is the case in the most profound examples. Expressing $U(t)=e^{-it\mathcal{H}}$ in terms of $\mathcal{H}$ is equivalent to solving the Cauchy problem (\ref{CPgen}) for each $\psi_0\in L_2(Q)$, and usually Schr\"odinger equation (\ref{CPgen}) is a partial differential equation which is difficult to solve. There are several known cases when the hamiltonian $\mathcal{H}$ of the system is so simple that the solution of the Cauchy problem (\ref{CPgen}) is expressible via one simple formula, e.g. when we deal with quantum harmonic oscillator. But in general case such formulas are unknown. 

On the other hand, if we succeed in finding a strongly continuous family of bounded self-adjoint operators that are Chernoff-tangent (see definition \ref{CTdef}) to the operator $\mathcal{H}$, then we can apply theorem \ref{RemQ} which allows to obtain $U(t)$ and $\psi(t)$ in the form of an expression that includes multiple integrals of arbitrary high  miltiplicity and Dirac $\delta$-functions under the integral sign (see subsection \ref{qFFsubs}). In the present paper we obtain such an expression for when (\ref{CPgen}) is representable in the form (\ref{CP1}) or (\ref{f_Koshi}).

For physical applications one often needs to calculate so-called matrix elements  $\left<U(t)\psi_1,\psi_2\right>$ for some $\psi_1,\psi_2\in L_2(Q)$. This problem is easier to solve if we have a formula for $U(t)$ which is more useful than $U(t)=\exp(-it\mathcal{H})$ --- this one is just a way to express that $-i\mathcal{H}$ is an infinitesimal generator of $C_0$-group $\left(\exp(-it\mathcal{H})\right)_{t\in\mathbb{R}}$, but not a way of calculating $U(t)$. Because of the quantum mechanical significance,
the properties of $\exp(-it\mathcal{H})$ have been extensively studied. Research topics include: exact solutions to the Cauchy problem, asymptotic behavior, estimates, related spatio-temporal structures, wave traveling, boundary conditions, etc. Some of the recent papers related to solution of the Cauchy problem for the Scr\"odinger equation are \cite{HW, Nak, Ord,  WZ, CorN, CNR,IN, Maz,Ah,HMmung}, see also \cite{Dobretal}.

\subsection{Feynman formulas and Quasi-Feynman formulas}\label{qFFsubs}  \textit{Feynman formula} (in sence of Smolyanov \cite{STT}) is an equality of the following form: on the left-hand side we have a function defined by the equality, and on the right-hand side we have a limit of multiple integral where the miltiplicity tends to infinity. Suppose that function $u(t,x)$ is the solution for the following Cauchy problem: $u'_t=Lu, u(0,x)=u_0(x)$. The expression $$u(t,x)=\lim_{n\to\infty}\underbrace{\int_E\dots\int_E}_{n}\dots dx_1\dots dx_n$$
is called \textit{a Lagrangian Feynman formula} if $E$ is a configuration space for the dynamical system that is described by the equation $u'_t=Lu$; it is called \textit{a Hamiltonian Feynman formula} if $E$ is a phase space for the same system. For the first time Lagrangian Feynman formulas appeared in the paper by R.\,P.~Feynman \cite{F1} in 1948, who postulated them without proof. The proof based on the Trotter product formula was provided by E.~Nelson \cite{Nel} in 1964. Hamiltonian Feynman formulas were presented in Feynman's paper \cite{F2} in 1951, but the proof (based on the Chernoff theorem) was published only in 2002 by O.\,G.~Smolyanov, A.\,G.~Tokarev and A.~Truman \cite{STT}. Pre-limit expressions in Feynman formulas approximate Feynman path integrals, which can be seen in \cite{STT, AHKM} and references therein.

Since 2000, O.\,G.~Smolyanov and members of his group succeeded in representing solutions of the Cauchy problem for many evolution equations in form of Feynman formulas (see \cite{Plya1, Plya2, Butko1, Butko2, OSS, R2, SWW, BGS2010,Dubravina, Bmnog, SWW,SYaa, SSWW, SakTMF, Buz2017, R8} and refereces therein). The key idea in these representations lies in finding the Chernoff function $G$ for operator $L$ and then applying Chernoff's theorem to obtain the equality $$e^{tL}u_0=\lim_{n\to\infty}G(t/n)^nu_0$$
which apperas to be a Feynman formula, because in all known examples (until \cite{R3} was published in 2016, see also \cite{R8, R-AMC}) $G(t)$ from the equation above was an integral operator, so $G(t/n)^n$ was an $n$-tuple integral operator, giving us a limit of multiple integral where miltiplicity tends to infinity.

For the case of Schr\"odinger equation ($L=iH$, where $H$ is a self-adjoint operator equal to  Hamiltonian with inverse sign, $H=-\mathcal{H}$) another approach was proposed in 2014 \cite{R4} (published with full proof in 2016 \cite{R1}). Proposed idea is as follows: we find $S$ that is Chernoff-tangent to $H$ (e.g. $S$ is a Chernoff function for $H$ if we know it or $S(t)=e^{tH}$) and then construct the Chernoff function for $iH$ via the formula $R(t)=e^{i(S(t)-I)}$, where $I$ is the identity operator.  There are no problems defining the exponent because for each $t$ operator $i(S(t)-I)$ is bounded. All conditions (CT) for $R$ follow from (CT) for $S$. And if we have chosen $S$ in such a way that it is self-adjoint ($S(t)^*=S(t)$), then  operator $A=S(t)-I$ is also self adjoint, and we have $\|R(t)\|=\left\|e^{iA}\right\|=1$ as a corollary from the Stone's theorem, so (N) for $R$ is satisfied. Formal statement follows.

\begin{theorem}
\label{RemQ}(\textsc{I.\,D.~Remizov, 2016}; new wording of theorem 3.1 from \cite{R1}).  Let $\mathcal{F}$ be a complex Hilbert space and let $Dom(H)\subset\mathcal{F}$ be its dense linear subspace. Suppose that operator $H\colon Dom(H) \to \mathcal{F}$ is linear and self-adjoint, and real number $a$ is nonzero. Suppose that we have such a family $(W(t))_{t\geq 0}$ of bounded linear operators in $\mathcal{F}$ that $(W(t))^*=W(t)$ for each $t \geq 0$, and, denoting $S(t)=I+W(t)$, the family $(S(t))_{t\geq 0}$ is Chernoff-tangent to $H$. Set $R(t) =\exp\big[ia(S(t)-I)\big]= \exp\big[iaW(t)\big].$ (This expression is well-defined because for each  $t\geq 0$ in the power of exponent only linear bounded operators in $\mathcal{F}$ appear.)

Then there exists a $C_0$-semigroup $\left(e^{iatH}\right)_{t \geq 0}$, family $(R(t))_{t\geq0}$ is Chernoff-equivalent to this semigroup, and for each $f \in \mathcal{F}$ and each $t_0\geq 0$ the following equalities hold with respect to norm in $\mathcal{F}$: $$e^{iatH}f=\lim_{n\to+\infty}R(t/n)^nf=\lim_{n\to+\infty}\exp\big[ianW(t/n)\big]f,\quad 0\leq t\leq t_0,$$
\begin{equation}\label{f_r2}
\lim_{n\to+\infty}\sup_{t\in[0,t_0]}\left\|e^{iatH}f-\lim_{j\to+\infty}\sum_{k=0}^{j}\frac{(ian)^k}{k!}W(t/n)^kf\right\|=0.\end{equation}
\end{theorem}

\begin{remark}\label{remnorm}
In short, theorem states the following: if $S$ is Chernoff tangent to $H$, and operators $H$ and $S(t)$ are self-adjoint, then $R(t)=e^{i(S(t)-I)}$ is Chernoff-equivalent to $(e^{itH})_{t \geq 0}$. The difference between $S$ and $W$ is that $S(0)=I$ and $W(0)=0$ which makes expression (\ref{f_r2}) simpler than the original form in theorem 3.1 in \cite{R1}. Note that we have NOT used the norm bound condition (N) for $S$ here, but still achieved (N) for $R$, so this approach is more flexible than the standard procedure of finding a family of integral operators that is Chernoff-equivalent to $C_0$-semigroup $\left(e^{itH}\right)_{t\geq 0}$. This flexibility will be highly used in the present paper. Indeed, if we set $A = S(t) - I$ in the Stone theorem 1.1, we get that $A$ is self-adjoint and $\left\|e^{i(S(t) -I)}\right\| = \left\|e^{iA}\right\| = 1$ because $e^{iA}$ is unitary thanks to the Stone theorem. We can also set $R(t)=e^{ia(S(t)-I)}$ for each nonzero number $a\in\mathbb{R}$, and this family will be Chernoff equavalent to $C_0$-semigroup $\left(e^{iatH}\right)_{t\geq 0}$. One can consider $a=1$ or $a=-1$ and study "forward" and "back" evolution. Generalization of this idea can be found in \cite{R5}.
\end{remark}

\textit{Quasi-Feynman formula} (in sence of \cite{R1}) is an equality of the following form: on the left-hand side we have a function defined by the equality, and on the right-hand side we have an expression that includes multiple integrals of arbitrary high  miltiplicity. The difference from a Feynman formula is that a quasi-Feynman formula may include summation or other operations on multiple integrals on the right-hand side, while only one multiple integral is allowed in a Feynman formula. If  $W(t)$ is an integral operator, then (\ref{f_r2}) is a quasi-Feynman formula.  

Quasi-Feynman formulas are lengthier than Feynman formulas but easier to obtain. Also, construction of Chernoff functions to solve Scr\"o\-din\-ger equation $\psi'(t)=iH\psi(t)$ is more difficult that doing the same for equation $\psi'(t)=H\psi(t)$. Let us provide several examples.

A.S. Plyashechnik in 2012-2013 obtained \cite{Plya1,Plya2} Feynman formulas for heat equation and Schr\"odinger equation in $\mathbb{R}^n$ with time- and space- dependent coefficients; the case of Schr\"odinger equation took more effort --- it required regularization with small $\varepsilon>0$ which depends on $n$ and appears in the final Feynman formula. Feynman formulas for parabolic (heat-type) equation with variable coefficients in infinite-dimensional Hilbert space were obtained in 2012 \cite{R2}, and for corresponding Schr\"odinger the question of proving such formulas is still open (but see \cite{SShav}), meanwhile V.Zh. Sakbaev in 2017  \cite{SakTMF} constructed quasi-Feynman formulas for this equation using theorem \ref{RemQ}. M.S. Buzinov in 2015 has obtained \cite{Buz2015, Buz2017, R1} Feynman formulas for heat-type evolution equation with natural power of Laplacian on the right-hand side of the equation, but for corresponding Schr\"odinger equation he only constructed quasi-Feynman formulas using theorem \ref{RemQ}. See also  section 6 of \cite{OSS-2016} where authors provide solution for a particular case of Schr\"odinger equation with constant coefficients and derivative of 6-th order in the Hamiltonian. See also \cite{M1, M2} and references therein. 

In the present paper, we express solution of the Cauchy problem (\ref{CP1}) in terms of coefficients of (\ref{CP1}). We provide a family of translation operators that is Chernoff-tangent to self-adjoint operator from (\ref{CP1}) and then apply theorem \ref{RemQ}. Then we do the same for (\ref{f_Koshi}). We come to formulas that do not include integrals at all, but then interpret expressions obtained as quasi-Feynman formulas with Dirac $\delta$-functions under the integral sign. 

This approach was used first in \cite{R3} for a simple case of one-dimensional Schr\"odinger equation with the second derivative only and bounded potential in the Hamiltonian. In the present paper we develop methods of \cite{R3} in two directions. Firsly, we cover the case of the Hamiltonian with derivatives of higher order in one-dimensional case. Secondly, we consider a multi-dimensional space in the case when Hamiltonian has only two terms: the Laplacian and potential. In both cases the potential may be unbounded which covers the Hamiltonian of quantum (an)harmonic oscillator, this was not done in \cite{R3}.  See also \cite{R7} for short introduction to quasi-Feynman formulas and the  calculus of Chernoff functions.

\section{Heuristic arguments for one-dimensional equation}\label{heur-arg}
In this section we construct a formula to define a Chernoff function for the Sturm-Liouville operator, which allows us to obtain the solution to the Cauchy problem  for Schr\"odinger equation with the Sturm-Liouville operator. We do not prove the formula here, but show how one can come to the formula in this case or in similar cases: technical formalities often change from case to case, but the idea stands more or less the same, and we show this idea.  We also develop an idea that is applicable to the case of equations of higher order, allowing us to solve (\ref{CP1}). Formal statement and the proof are presented in the next section.

\subsection{Construction blocks}
Consider a smooth bounded function $p\colon\mathbb{R}\to\mathbb{R}$, a measurable unbounded function  $q\colon\mathbb{R}\to\mathbb{R}$, a smooth bounded function $w\colon\mathbb{R}\to\mathbb{R}$ with $w(0)=0$ and $w'(0)=1$ and a fixed number $t\in\mathbb{R}$, define the following bounded operators in complex  $L_2(\mathbb{R})$ (the star $^*$ is used to show that operator $Z^*$ is adjoint to operator $Z$):
$$
(B_pf)(x)=(B_p^*f)(x)=p(x)f(x),
$$ 
$$
(B_{wq}(t)f)(x)=(B_{wq}(t)^*f)(x)=w(tq(x))f(x),
$$ 
$$
(A(t)f)(x)=f(x+t),\quad (A(t)^*f)(x)=f(x-t)
$$
and the following unbouded operators:
$$
(B_{q}f)(x)=(B_{q}^*f)(x)=q(x)f(x),\quad\textrm{(multiplication by } q\textrm{),}
$$
$$
(\partial f)(x)=f'(x) \quad\textrm{ (differentiation),}
$$
$$
(\partial B_p\partial +B_q) f(x)=(p(x)f'(x))' +q(x)f(x) \quad\textrm{ (Sturm-Liouville operator).}
$$
We assume that functions $p$ and $q$ have been chosen in such a way that the Sturm-Liouville operator is defined on some dense linear subspace of complex $L^2(\mathbb{R})$ and is self-adjoint.

\subsection{Sturm-Liouville operator, zero potential}
Let us first consider a simple case of $q(x)\equiv 0$, then the Cauchy problem for Schr\"odinger equation with the Sturm-Liouville operator reads as
$$
\left\{ \begin{array}{ll}
\psi'_t(t,x)=i\partial B_p\partial \psi(t,x)\\ 
\psi(0,x)=\psi_0(x)
\end{array} \right.
$$
and is known to have the solution $$\psi(t,x)=\exp[it\partial B_p\partial]\psi_0.$$
The only problem is that we cannot calculate the bounded operator  $\exp[it\partial B_p\partial]$ directly from this formula because we have an unbounded operator in the power of the exponent, making the power series $e^Z=\sum_{n=0}^\infty Z^n/n!$ useless to us. However, we can apply the approach based on Chernoff tangency and theorem \ref{RemQ}.

It is known (and also not difficult to show by checking the conditions of definition~\ref{semigrdef}) that $\left(A(t)\right)_{t\in\mathbb{R}}$ and $\left(A(t)^*\right)_{t\in\mathbb{R}}$ are $C_0$-groups in $L_2(\mathbb{R})$. The infinitesimal generators of those groups are $\partial$ and $-\partial$ respectively, which implies that
$$A(t)f=f + t\partial f+o(t),\quad A(t)^*f=f - t\partial f+o(t).$$
So $A(t)$ is Chernoff-tangent to $\partial$, $A(t)^*$ is Chernoff-tangent to $-\partial$, and we need to somehow combine them with $B_p$ to yield such $S_1(t)$ that $S_1(t)=S_1(t)^*$ and $S_1$ is Chernoff-tangent to $\partial B_p\partial$. We can see that we need to obtain 
$S_1(t)f=f +t\partial B_p\partial f +o(t)$ from conditions (CT2) and (CT3). We write $S_1$ instead of $S$ in theorem \ref{RemQ} for the reason that will be clear below. One of the possible formulas for $S_1(t)$ is 
$$
S_1(t)=F_1(t)+I,
$$ 
where  
$$
F_1(t)=\left(A(\sqrt{t})-I\right)B_p\left(I-A(\sqrt{t})^*\right).
$$
Let us show that $S_1(t)=S_1(t)^*$. Indeed, $S_1(t)^*=(F_1(t)+I)^*=F_1(t)^*+I$ so it is enough to show that $F_1(t)^*=F_1(t)$. We have
$F_1(t)^*=\Big(\left(A(\sqrt{t})-I\right)B_p$ $\left(I-A(\sqrt{t})^*\right)\Big)^*=\Big(B_p\left(I-A(\sqrt{t})^*\right)\Big)^*$ $\Big(A(\sqrt{t})-I\Big)^*=\left(I-A(\sqrt{t})\right)B_p^*$ $\left(A(\sqrt{t})^*-I\right)=\left(A(\sqrt{t})-I\right)B_p\big(I-  
A(\sqrt{t})^*\big)=F_1(t)$. 

Let us see what happens when $t$ tends to zero: 
$
F_1(t)=\left(A(\sqrt{t})-I\right)B_p$ $\left(I-A(\sqrt{t})^*\right)=\big(I+\sqrt{t}\partial$  $+o(t)-I\big)B_p \big(I-(I-$ $\sqrt{t}\partial +o(t))\big)=t\partial B_p\partial +o(t).$ Hence we have
$$S_1(t)=I+F_1(t)=I+t\partial B_p\partial +o(t).$$

Now we can define $R_1(t)=\exp[i(S_1(t)-I)]=\exp[iF_1(t)]$ which implies (by theorem \ref{RemQ}) that $R_1(t)=I+it\partial B_p\partial +o(t)$ and $\exp[it\partial B_p\partial]=\lim_{n\to\infty} R_1(t/n)^n$. 

\subsection{Sturm-Liouville operator, nonzero potential}
Let us go back to the general case $q(x)\not\equiv 0$. We now deal with the Cauchy problem for Schr\"odinger equation with the Sturm-Liouville operator 
$$
\left\{ \begin{array}{ll}
\psi'_t(t,x)=i(\partial B_p\partial +B_q) \psi(t,x)\\ 
\psi(0,x)=\psi_0(x)
\end{array} \right.
$$
and need to find a formula for the solution $$\psi(t,x)=\exp[it(\partial B_p\partial + B_q)]\psi_0.$$

First idea that comes to mind is to use the famous \cite{EN1} Trotter's product formula $e^{X+Y}=\lim_{n\to\infty}\left(e^{X/n}e^{Y/n}\right)^n$, but this will lead us to a triple limit expression (two limits from theorem \ref{RemQ} and one from the Trotter's formula). To avoid this we will modify the above constructed family $S_1(t)$ by somehow increasing the derivative at zero by $B_q$ and only after this apply theorem \ref{RemQ}.

Another challenge lies in that function $q$ is not bounded, so the operator $B_q$ is also not bounded, making the operator-valued
function $S_{\textrm{wrong}}(t)=I+F_1(t)+tB_q$ not Chernoff-tangent to $\partial B_p\partial+B_q$ because operator $S_{\textrm{wrong}}(t)$ becomes unbounded which contradicts (CT0). To overcome this we will multiply $f(x)$ not by $tq(x)$, but by a bounded function $w(tq(x))$, where 
$w(0)=0$ and $w'(0)=1$. Indeed, operators  $F_0(t)=B_{wq}(t)$ are bounded and have the correct derivative at zero: $B_{wq}(t)f(x)=w(tq(x))f(x)=w(0)f(x)+ tw'(0)q(x)f(x)+o(t)=q(x)f(x)+o(t)=tB_qf(x)+o(t).$

Keeping all that in mind, we define $$S(t)=I+F_1(t)+F_0(t)=I+\left(A(\sqrt{t})-I\right)B_p\left(I-A(\sqrt{t})^*\right)+B_{wq}(t).$$ Operators $I$, $F_1(t)$ and $F_0(t)$ are bounded and self-adjoint, so their sum has the same properties. The derivative at zero is exactly the one we need:
$$
S(t)=I+F_1(t)+F_0(t)=I+t\partial B_p\partial +o(t) + tB_q+o(t)=I+ t(\partial B_p\partial+B_q)+o(t).
$$
Finally, by defining $R(t)=\exp[i(S(t)-I)]=\exp[i(F_1(t)+F_0(t))]$ and applying theorem \ref{RemQ} to obtain $R(t)=I+it(\partial B_p\partial+B_q) +o(t)$, we have $$\exp[it(\partial B_p\partial+B_q)]=\lim_{n\to\infty} R(t/n)^n.$$

\subsection{Operators of higher order}
The same technique works with fixed $k\in\mathbb{N}$: assume that function  $a_k\colon\mathbb{R}\to\mathbb{R}$ is measurable and bounded and replace the operator $\partial B_p \partial$ from above subsections with $\partial^k B_{a_k} \partial^kf(x)=\frac{d^k}{dx^k}\left(a_k(x)\frac{d^k}{dx^k}f(x)\right)$ . The corresponding family is $$S_k(t)=I+F_k(t),$$ where 
$$
F_k(t)=\left(A(t^{1/2k})-I\right)^kB_{a_k}\left(I-A(t^{1/2k})^*\right)^k.$$
Let us examine the behevoiur of this expression with $t$ tending to zero:
$$ F_k(t)=\left(t^{1/2k}\partial+o(t^{1/2k})\right)^kB_{a_k}\left(t^{1/2k}\partial+o(t^{1/2k})\right)^k,$$
$$F_k(t)=\left(t^{1/2}\partial^k+o(t^{1/2})\right)B_{a_k}\left(t^{1/2}\partial^k+o(t^{1/2})\right),
$$
$$
F_k(t)=t\partial^kB_{a_k}\partial^k+o(t).
$$

Let us define $\partial^0 B_{a_0} \partial^0f(x)=B_{a_0}f(x)=a_0(x)f(x)$ and  $F_0(t)f(x)=B_{wa_0}(t)f(x)=w(ta_0(x))f(x)$ to cover the case $k=0$. Then for each $k=0,1,2,\dots$ we have $F_k(t)^*=F_k(t)$ and $F_k(t)=t\partial^k B_{a_k} \partial^k +o(t).$
Now consider an operator
$$\mathcal{H}=\sum_{k=0}^K\partial^k B_{a_k} \partial^k$$
and define 
$
S(t)=I+\sum_{k=0}^K F_k(t).
$
Then $S(t)=S(t)^*$ and $S(t)=I +t\mathcal{H} +o(t)$. Note that we should not expect $\|S(t)\|\leq 1+\alpha t$ here, but this is not a problem due to remark \ref{remnorm}. With definitions of this subsection the Cauchy problem (\ref{CP1}) reads as
$$
\left\{ \begin{array}{ll}
\psi'_t(t)=-i\mathcal{H}\psi(t),\\ 
\psi(0)=\psi_0.
\end{array} \right.
$$
Applying theorem \ref{RemQ} with $a=-1$ we come to a formula $$R(t)=\exp[-i(S(t)-I)]=\exp\left[-i\sum_{k=0}^K F_k(t)\right]$$
and obtain the solution of (\ref{CP1}) in the form
$$
\psi(t,x)=\left(e^{tL}\psi_0\right)(x)=\left(\lim_{n\to\infty}R(t/n)^n\psi_0\right)(x).
$$

Now, having found the right formula for $S(t)$, let us state and prove theorem based on it.

\section{Main result}
Theorem statements and proofs in this section are intentionally made a bit wordy because we would like to keep them  self-contained in sense of notation and facts to help those who wish to skip the prelude and dig straight into the main result. However, all the symbols are in line with those provided in previous sections to help reader with connecting physical meaning with heuristic arguments and formal statements that will follow.

\subsection{One-dimensional Schr\"odinger equation}
\begin{theorem}\label{mainth}
Fix arbitrary $K\in\mathbb{N}$. Suppose that for $k=0,1,\dots,K$ functions  $a_k\colon\mathbb{R}\to\mathbb{R}$ are given. Suppose that for each $k=1,\dots,K$ function $a_k$ belongs to space $C_b^{2k}(\mathbb{R})$ of all bounded functions $\mathbb{R}\to\mathbb{R}$ with bounded derivatives up to $(2k)$-th order. Suppose that function $a_0\colon \mathbb{R}\to\mathbb{R}$ is measurable and belongs to space $L_2^{loc}(\mathbb{R})$, i.e. $\int_{-R}^{R}|a_0(x)|^2dx<\infty$ for each real number $R>0$. Define
$$
(\mathcal{H}\varphi)(x)=a_0(x)\varphi(x)+\sum\limits_{k=1}^{K}\frac{d^k}{dx^k}\left(a_k(x) \frac{d^k}{d x^k} \varphi(x)\right)
$$
for each $\varphi$ from the space $C_0^\infty(\mathbb{R})$ of all functions $\varphi\colon\mathbb{R}\to\mathbb{R}$ wich are bounded together with their derivatives of all orders and have compact support (are zero outside of some closed interval).
We also use the following condition for coefficients $a_k$, $k=0,1,\dots,K$: operator $\mathcal{H}$ defined on $C_0^\infty(\mathbb{R})$ is essentially self-adjoint in $L_2(\mathbb{R})$, i.e. the operator $(\mathcal{H}, C_0^\infty(\mathbb{R}))$ is closable and its closure --- let us denote it as $(\mathcal{H},Dom(\mathcal{H}))$ --- is a self-adjoint operator. 

Suppose that function $w\colon\mathbb{R}\to\mathbb{R}$ is continuous, bounded, differentiable at zero and  $w(0)=0$, $w'(0)=1$ (examples include: $w(x)=\arctan(x)$, $w(x)=\sin(x)$, $w(x)=\tanh(x)=(e^x-e^{-x})/(e^x+e^{-x})$, etc). For each $t\geq 0$, $k=1,2,\dots,K$, each $x\in\mathbb{R}$, and each $f\in L_2(\mathbb{R})$ define:
$$
(B_{a_k}f)(x)=a_k(x)f(x),
$$ 
$$
(A(t)f)(x)=f(x+t),\quad (A(t)^*f)(x)=f(x-t),
$$
$$
F_k(t)=\left(A(t^{1/2k})-I\right)^kB_{a_k}\left(I-A(t^{1/2k})^*\right)^k,\ \ F_0(t)f(x)=w(ta_0(x))f(x),
$$
\begin{equation}\label{FSdef}
F(t)=\sum_{k=0}^K F_k(t),\ \ S(t)=I+F(t)=I+\sum_{k=0}^K F_k(t),
\end{equation}
where $I$ is the identity operator ($If=f$), and expression such as $Z^k$ means the composition $ZZ\dots Z$ of $k$ copies of linear bounded operator $Z$. 

Then the following holds:

1) For each $t\geq 0$ operators $A(t)$, $A(t)^*$, $B_{a_k}$ for $k=1,2,\dots,K$, $F_k(t)$ for $k=0,1,\dots,K$ and $F(t)$, $S(t)$ are linear bounded operators in $L_2(\mathbb{R})$, and their norms are bounded by a constant that does not depend on $t$

2) $S$ is Chernoff-tangent to $\mathcal{H}$

3) $S(t)=S(t)^*$ for each $t\geq 0$

4) For each $t\geq 0$ operator  $R(t)=\exp[-iF(t)]$ is a well-defined linear operator in $L_2(\mathbb{R})$

5) There exists a $C_0$-group $\left(e^{-it\mathcal{H}}\right)_{t\in\mathbb{R}}$ of linear boounded unitary operators in $L_2(\mathbb{R})$

6) $R$ is Chernoff-equivalent to $\left(e^{-it\mathcal{H}}\right)_{t\in\mathbb{R}}$, and the following formulas hold for each $f\in L_2(\mathbb{R})$ and $t\geq 0$, where limits exist with respect to norm in $L_2(\mathbb{R})$: 
$$
e^{-it\mathcal{H}}=\lim_{n\to\infty}R(t/n)^n=\lim_{n\to\infty}\exp\left[-in F(t/n)\right]=\lim_{n\to\infty}\exp\left[-in\sum_{k=0}^K F_k(t/n)\right],
$$
$$
e^{-it\mathcal{H}}=\lim_{n\to\infty}\lim_{j\to+\infty}\sum_{q=0}^{j}\frac{(-in)^q}{q!}\left(\sum_{k=0}^K F_k(t/n)\right)^q. 
$$

7) For each initial condition $\psi_0\in L_2(\mathbb{R})$ the Cauchy problem (1) can be written in the form
$$
\left\{ \begin{array}{ll}
\psi'_t(t)=-i\mathcal{H}\psi(t),\\ 
\psi(0)=\psi_0,
\end{array} \right.
$$
and has a unique (in sense of $L_2(\mathbb{R})$) solution $\psi(t)$ that depends on $\psi_0$ continuously with respect to norm in $L_2(\mathbb{R})$, and for all $t\geq 0$ and almoust all $x\in\mathbb{R}$ can be expressed in the form
$$
\psi(t,x)=\left(e^{-it\mathcal{H}}\psi_0 \right)(x)=\left(\lim_{n\to\infty}\lim_{j\to+\infty}\sum_{q=0}^{j}\frac{(-in)^q}{q!}\left(\sum_{k=0}^K F_k(t/n)\right)^q\psi_0\right)(x).
$$
Here linear bounded operators $F_0(t),\dots, F_K(t)$ are defined above in conditions of the theorem for all $t\geq 0$ (hence $F_0(t/n),\dots, F_K(t/n)$ are defined for all $t\geq 0$ and all $n\in\mathbb{N}$), and the power $q$ in $\left(\sum_{k=0}^K F_k(t/n)\right)^q$ stands for a composition of $q$ copies of linear bounded operator  $\sum_{k=0}^K F_k(t/n)$.
\end{theorem}
\textbf{Proof.} The structure of the proof is the following. We derive items 1)-3) from conditions of the theorem, and see that item 4) follows from item 1). After that we apply Stone's theorem (theorem \ref{Stth}) to get item 5) and theorem \ref{RemQ} to get item 6). Item 7) then follows from item 6) and general facts of $C_0$-semigrops theory that are listed in subsection \ref{semgreveq}.

Item 1). Recall that for $k=1,\dots,K$ function $a_k$ is bounded, so 
$\|B_{a_k}f\|=$  $(\int_{\mathbb{R}}|a_k(x)f(x)|^2dx)^{1/2}\leq (\sup_{x\in\mathbb{R}}|a_k(x)|^2\int_{\mathbb{R}}|f(x)|^2dx)^{1/2}=\|f\|\sup_{x\in\mathbb{R}}|a_k(x)|$, which implies $\|B_{a_k}\|\leq \sup_{x\in\mathbb{R}}|a_k(x)|<\infty$. Function $a_0$ is not bounded, but fuction $w$ is bounded, hence function $x\longmapsto w(ta_0(x))$ is bounded and we can estimate $\|F_0\|$ in the same manner as above: $\|F_0(t)f\|\leq \|f\|\sup_{x\in\mathbb{R}}|w(ta_0(x))|$, so $\|F_0(t)\|\leq \sup_{z\in\mathbb{R}}|w(x)|\equiv\mathrm{const}<\infty$ for all $t\geq 0$. Change of variable $y=x+t, dy=dx$ in the integral $\|A(t)f\|=(\int_{\mathbb{R}}|f(x+t)|^2dx)^{1/2}=(\int_{\mathbb{R}}|f(y)|^2dy)^{1/2}=\|f\|$ shows that $\|A(t)\|=1$ for all $t\geq 0$, and similarly $\|A(t)^*\|=1$ for all $t\geq 0$. Operator $F_k(t)$ is obtained via finite number of summations and compositions of bounded operators whose norm is bounded by a constant that does not depend on $t$, so $F_k(t)$ has the same property. Then $F(t)$ and $S(t)$ also have this property.

Item 2). In definition \ref{CTdef} we set $\mathcal{F}=L_2(\mathbb{R})$, $G(t)=S(t)$, $L=\mathcal{H}$, $\mathcal{D}=C_0^\infty(\mathbb{R})\subset L_2(\mathbb{R})$. We do not have the precise description of $Dom(L)\subset L_2(\mathbb{R})$, but we say that $Dom(L)$ is the domain of the closure of the operator $\mathcal{H}$ on the domain $C_0^\infty(\mathbb{R})$; by conditions of the theorem this closure exists and can be shown to be a self-adjoint operator in $L_2(\mathbb{R})$. Now let us check (CT) for $S$ and $\mathcal{H}$.

(CT0) follows from the prelude above and item 1) which states that for each $t\geq 0$ we have $S(t)\in \mathscr{L}( L_2(\mathbb{R}))$.

(CT1) We need to prove that for each fixed $f\in L_2(\mathbb{R})$ the mapping $t\longmapsto S(t)f\in L_2(\mathbb{R})$ is continuous. Given $t_0\geq0$ and $t_n\geq0$ with $t_n\to t_0$ we need to show that $\lim_{n\to\infty}\|S(t_n)f-S(t_0)f\|=0$. We will do it in four steps i)-iv).

i). Let us first show that $\|F_0(t_n)f-F_0(t_0)f\|\to0$. Indeed, $\|F_0(t_n)f-F_0(t_0)f\|^2=\int_{\mathbb{R}}|w(t_na_0(x))-w(t_0a_0(x))|^2|f(x)|^2dx$. As $w$ is continuous, for each $x\in\mathbb{R}$ we have $w(t_na_0(x))\to w(t_0a_0(x))$, so the integrand in the above integral converges to zero pointwise. As $f\in L_2(\mathbb{R})$ and function $w$ is bounded we can apply the Lebesgue dominated convergence theorem and be sure that $\int_{\mathbb{R}}|w(t_na_0(x))-w(t_0a_0(x))|^2|f(x)|^2dx\to0$. So $\|F_0(t_n)f-F_0(t_0)f\|^2\to0$ which implies $\|F_0(t_n)f-F_0(t_0)f\|\to0$.

ii). Let us show that $\|F_k(t_n)f-F_k(t_0)f\|\to0$ for each fixed $k=1,\dots,K$. We reduce this task to a simpler one. If we expand the brackets in the equality
$$F_k(t)=\left(A(t^{1/2k})-I\right)^kB_{a_k}\left(I-A(t^{1/2k})^*\right)^k,$$
we will see that $F_k(t)f$ is a finite sum of elements of the form $$(-1)^{j_1}A(t^{1/2k})^{j_2}B_{a_k}A(t^{1/2k})^{*j_3}f,$$ where $j_1,j_2,j_3$ are some nonnegative integers. So to show that $t\longmapsto F_k(t)f$ is continuous it is enough to show that $t\longmapsto A(t^{1/2k})^{j_2}B_{a_k}A(t^{1/2k})^{*j_3}f$ is continuous for each integer $j_2\geq 0, j_3\geq 0$. By definition $(A(t)f)(x)=f(x+t)$ and $(A(t)^*f)(x)=f(x-t)$, so $(A(t^{1/2k})^{j_2}f)(x)=f(x+j_2t^{1/2k})$ and $A(t^{1/2k})^{*j_3}f=f(x-j_3t^{1/2k})$. Recalling that $(B_{a_k}f)(x)=a_k(x)f(x)$ we come to the following formula:
\begin{align*}
\left(A(t^{1/2k})^{j_2}B_{a_k}A(t^{1/2k})^{*j_3}f\right)(x) &\stackrel{\mathrm{~~~~~~}}{=}a_k(x+j_2t^{1/2k})f(x+(j_2-j_3)t^{1/2k})\\
&\stackrel{\mathrm{denote}}{=}m(t)f(x).
\end{align*}
So\\   
$\|m(t_0)f-m(t_n)f\|^2=$ $ \|A(t_0^{1/2k})^{j_2}B_{a_k}A(t_0^{1/2k})^{*j_3}f - A(t_n^{1/2k})^{j_2}B_{a_k}A(t_n^{1/2k})^{*j_3}f\|^2$ $=\int_{\mathbb{R}}|a_k(x+j_2t_0^{1/2k})f(x+(j_2-j_3)t_0^{1/2k}) -a_k(x+j_2t_n^{1/2k})f(x+(j_2-j_3)t_n^{1/2k})|^2dx$. Function $a_k$ in the last integral is bounded and continuous but $f$ is not, so we should not expect the integrand to tend to zero pointwise and can not apply Lebesgue theorem as easily as in step i). Instead, we will use the fact that $C_0^\infty(\mathbb{R})$ is dense in $L_2(\mathbb{R})$ and apply the so-called  "$\varepsilon/3$-method" in step iii). 

iii). We want to show that for arbitrary fixed $\varepsilon>0$ there exists $n_0\in\mathbb{N}$ such that $\|m(t_0)f-m(t_n)f\|<\varepsilon$ for all $n>n_0$. We have shown in item 1) that there exists a constant such that $\max_{k = 1 \ldots K} \sup_{t \geq 0} \|F_k(t)\| < \infty$. So $\|m(t)\|\leq M$ for some fixed  $M\in\mathbb{R}$ and all $t\geq 0$. As $C_0^\infty(\mathbb{R})$ is dense in $L_2(\mathbb{R})$, there exists such $g\in C_0^\infty(\mathbb{R})$ that $\|f-g\|<\varepsilon/(3M)$. Then
$$
\|m(t_0)f-m(t_n)f\|=\|m(t_0)f -m(t_0)g+m(t_0)g-m(t_n)g+m(t_n)g-m(t_n)f\|
$$
$$
\leq\|m(t_0)f -m(t_0)g\|+\|m(t_0)g-m(t_n)g\|+\|m(t_n)g-m(t_n)f\|
$$
$$
\leq\|m(t_0)\|\cdot \|f -g\|+\|m(t_0)g-m(t_n)g\|+\|m(t_n)\|\cdot \|f -g\|
$$
$$
< M\frac{\varepsilon}{3M} +\|m(t_0)g-m(t_n)g\|+M\frac{\varepsilon}{3M}.
$$

Now recall that functions $g$ and $a_k$ are continuous, so integrand in
$\|m(t_0)g-m(t_n)g\|^2=\int_{\mathbb{R}}|a_k(x+j_2t_0^{1/2k})g(x+(j_2-j_3)t_0^{1/2k}) -a_k(x+j_2t_n^{1/2k})g(x+(j_2-j_3)t_n^{1/2k})|^2dx$ converges to zero pointwise (for each $x\in\mathbb{N}$) as $n\to\infty$. Function $a_k$ is bounded, and $|g|^2$ is integrable (recall that $g$ is zero everywhere outside some closed interval), so we can apply the Lebesgue dominated convergence theorem and obtain $\lim_{n\to\infty}\|m(t_0)g-m(t_n)g\|=0$. Then there exists $n_0\in\mathbb{N}$ such that
for all $n>n_0$ we have $\|m(t_0)g-m(t_n)g\|<\varepsilon/3$. Combining this with the previous inequality we obtain
$$
\|m(t_0)f-m(t_n)f\|< M\frac{\varepsilon}{3M} +\frac{\varepsilon}{3}+M\frac{\varepsilon}{3M}=\varepsilon.$$

iv). In steps ii) and iii) we have shown that for arbitrary fixed $f\in L_2(\mathbb{R})$ the mapping $t\longmapsto F_k(t)f$ is continuous for $k=1,\dots,K$, and in step i) that it is continuous for $k=0$. So finite sums $t\longmapsto\sum_{k=0}^K F_k(t)f=F(t)f$ and $t\longmapsto (F(t)f+f)=S(t)f$ also define continuous mappings. Now (CT1) is proven.

(CT2) follows directly from formula (\ref{FSdef}) and formulas above it. If we assume $t=0$ in $(A(t)f)(x)=f(x+t)$ we see that $A(0)=I$. The same simple check shows that $A(0)^*=I$ and $F_0(0)=I$. So $F_k(0)=0$ and $S(0)=I+F(0)=I+0=I$.

(CT3) is the most complicated part of the whole proof. Due to technical complexity of the reasoning that will follow we recommend reading the second section (which presents heuristic arguments) before the proof of (CT3) in order to keep the main idea in mind. However, the proof is self-contained so the reader may ignore this advice.

For each fixed $\varphi\in\mathcal{D}=C_0^\infty(\mathbb{R})$ we need to show that $S(t)\varphi=\varphi+t\mathcal {H}\varphi+o(t)$ as $t\to0$. Note that $o(t)$ is used in sence of $L_2(\mathbb{R})$, i.e. $\theta(t,x)=o(t)$ iff $\lim_{t\to0}t^{-1}\left(\int_\mathbb{R}|\theta(t,x)|^2dx\right)^{1/2}=0$. The proof is separated into eight steps i)-viii).

i). Recall that $S(t)\varphi=\varphi+F_0(t)\varphi+\sum_{k=1}^K F_k(t)\varphi$ and $(\mathcal{H}\varphi)(x)=a_0(x)\varphi(x)+\sum\limits_{k=1}^{K}\frac{d^k}{dx^k}\left(a_k(x) \frac{d^k}{d x^k} \varphi(x)\right)$. In step ii) we show that $F_0(t)\varphi=ta_0\varphi+o(t)$. In steps iii)-viii) we show that for $k=1,\dots,K$ we have 
$(F_k(t)\varphi)(x)=t\frac{d^k}{dx^k}\left(a_k(x) \frac{d^k}{d x^k} \varphi(x)\right)+o(t)$. Due to the just mentioned definitions of $S(t)$ and $\mathcal{H}$ this will be enough to reach our goal.

ii). Recall that function $w$ is bounded, continuous,  differentiable at zero and satisfies $w(0)=0$ and $w'(0)=1$. So, by Taylor's expansion formula with the remainder in Peano's form, $w$ can be represented as $$w(z)=z+zh(z),$$ where $\lim_{z\to0}h(z)=0$. Let us show that function $h$ is continuous and bounded. Let us define $h(0)=0$ and $h(z)=(w(z)-z)/z$ for $z\neq 0$. Function $w$ is continuous for all $z\in\mathbb{R}$, so $h$ is continuous for $z\neq 0$ due to the formula $h(z)=(w(z)-z)/z$, and $h$ is continuous at zero due to condition $\lim_{z\to0}h(z)=0=h(0)$. Now let us prove that $h$ is bounded.  Indeed, from $\lim_{z\to0}h(z)=0$ we get that $\sup_{|z|\leq 1}|h(z)|<\infty$. And for $|z|>1$ we can estimate  $|h(z)|=|w(z)/z - 1|\leq|w(z)/z| +1\leq |w(z)|+1<\infty$ because $w$ is bounded.

So for each $x\in\mathbb{R}$ and $z=ta_0(x)$ we have
$$
w(ta_0(x))\varphi(x)=ta_0(x)\varphi(x)+ta_0(x)\varphi(x)h(ta_0(x)).
$$ 

Now let us show that  $a_0(x)\varphi(x)h(t_na_0(x))\to0$ in $L_2(\mathbb{R})$ if $t_n\to 0$. Indeed, functions $\varphi$ and $h$ are bounded, and $a_0\in L^{loc}_2(\mathbb{R})$. Then functions $x\longmapsto|a_0(x)\varphi(x)h(t_na_0(x))|^2$ are: a) integrable on $[-R,R]$ ($\varphi$ is zero outside this segment); b)  majorated on this segment by an integrable function $x\longmapsto\left|a_0(x)\varphi(x)\sup_{z\in\mathbb{R}}|h(z)|\right|^2$; c) converging to zero for each $x\in [-R,R]$ as $n\to\infty$ because  $\lim_{z\to0}h(z)=0$, and $t_na_0(x)\to 0$. Then $\|a_0(\cdot)\varphi(\cdot)h(t_na_0(\cdot))\|^2=\int_{\mathbb{R}}|a_0(x)\varphi(x)h(t_na_0(x))|^2dx=\int_{-R}^R|a_0(x)\varphi(x)h(t_na_0(x))|^2dx\to0$ thanks to Lebesgue's dominated convergence theorem. As $\|\psi_n\|^2\to0$ implies $\|\psi_n\|\to0$, we conclude that $\lim_{n\to\infty}\|a_0(\cdot)\varphi(\cdot)h(t_na_0(\cdot))\|=0$. So we have proved that  
\begin{equation}\label{ocii}
(F_0(t)\varphi)(x)=w(ta_0(x))\varphi(x)=ta_0(x)\varphi(x)+o(t).
\end{equation}

iii). Let us say that $f\in C_0^p(\mathbb{R})$ iff function $f\colon \mathbb{R}\to\mathbb{R}$ is zero outside of some closed interval, is bounded, and has derivatives of orders $1,\dots,p$, which are also all bounded. Let us say that $f\in C_0^{ebd}(\mathbb{R})$ iff function $f\colon  \mathbb{R}\to\mathbb{R}$ is zero outside of some closed interval, is bounded, and has \textit{enough bounded derivatives} to make our reasoning (that will follow) true. Let us agree that the symbol $C_0^{ebd}(\mathbb{R})$ in different places may mean different spaces, similar to the agreement that allows us to use the same symbol $o(t)$ for different expressions in one formula. Example: the operator of differentiation maps $C_0^{ebd}(\mathbb{R})$ into $C_0^{ebd}(\mathbb{R})$, but maps $C_0^p(\mathbb{R})$ into $ C_0^{p-1}(\mathbb{R})$. We will use this agreement and prove some statements. After that we will go through the proof and find out how many bounded derivatives do we really need. This means that we find such $p\in\mathbb{N}$ that $C_0^{ebd}(\mathbb{R})=C_0^p(\mathbb{R})$ in the place where $C_0^{ebd}(\mathbb{R})$ denotes the smallest space of all $C_0^{ebd}(\mathbb{R})$ that we have actually used.

Fix arbitrary $f\in C_0^{ebd}(\mathbb{R})$, arbitrary $k=1,2,\dots$, arbitrary  $x\in\mathbb{R}$ and $\tau\in\mathbb{R}$, $\tau\neq 0$,  then by Taylor's formula with remainder in Lagrange's form we have
$$
f(x+\tau)=f(x)+\tau f'(x)+\dots+\frac{\tau^{k}}{k!}
 f^{(k)}(x)+\frac{\tau^{k+1}}{(k+1)!}
 f^{(k+1)}(\xi(x,\tau)),
$$
where number $\xi(x,\tau)$ is between $x$ and $x+\tau$. This equality can be rewritten as
\begin{equation}\label{razl}
(A(\tau)-I)f = \left(\tau\partial+\dots+\frac{(\tau\partial)^{k}}{k!}+\tau^{k+1}\Theta(\tau)\right)f,
\end{equation}
where we use the following notation $$(\Theta(\tau)f)(x)=\frac{1}{(k+1)!} f^{(k+1)}(\xi(x,\tau)), \quad (\partial f)(x)=f'(x).$$
Let us study properties of the operator $\Theta(\tau)$. 

iii-1. For all $\tau\in\mathbb{R}$ and $k\geq 1$ we have
$\Theta(\tau)(C_0^{k+1}(\mathbb{R}))\subset C_0^{1}(\mathbb{R})$, $\Theta(\tau)(C_0^{ebd}(\mathbb{R}))\subset C_0^{ebd}(\mathbb{R})$ and $\Theta(\tau)(C_0^\infty(\mathbb{R}))\subset C_0^\infty(\mathbb{R})$; moreover, operator $\Theta(\tau)$ is linear on these three domains, which can be seen by representing $\Theta(\tau)f$ as $$\Theta(\tau)f=\frac{1}{\tau^{k+1}}(A(\tau)-I-\tau\partial-\dots-(\tau\partial)^{k}/k!)f.$$

iii-2. For each fixed $f\in C_0^{ebd}(\mathbb{R})$ there exists such a constant $C(f)\in\mathbb{R}$ that
\begin{equation}\label{octhet}
\sup_{\tau\in\mathbb{R}}\|\Theta(\tau)f\|\leq C(f).
\end{equation}
This is true due to the fact that $f$ is zero outside of some closed interval (say, $[a,b]$) and has bounded $(k+1)$-th  derivative, so we can estimate
$
\|\Theta(\tau)f\|=\left(\int_a^b|f^{(k+1)}(\xi(x,\tau))/(k+1)!|^2dx\right)^{1/2}\leq \frac{(b-a)^{1/2}}{(k+1)!}\sup_{z\in\mathbb{R}}|f^{(k+1)}(z)|\stackrel{\mathrm{denote}}{=}C(f).
$

iv). Let us prove that for each $k=1,2,\dots$ and each $f\in C_0^{2k}(\mathbb{R})$ we~have
\begin{equation}\label{shagleft} 
(A(\tau)-I)^kf=\tau^k\partial^kf+\tau^k\Phi(\tau,f),
\end{equation}
where $\Phi(\tau,f)\to 0$ in $L_2(\mathbb{R})$, i.e.
$\lim_{\tau\to0}\|\Phi(\tau,f)\|=0$, which we also denote as $\tau^k\Phi(\tau,f)=o(\tau^k)$. Let us prove (\ref{shagleft}) by induction on~$k$, supposing that $f\in C_0^{ebd}(\mathbb{R})$ and only in the last step c-5)  specifying that $f\in C_0^{2k}(\mathbb{R})$.

a) For $k=1$ the desired equality (\ref{shagleft}) reads as $(A(\tau)-I)f=\tau\partial f+\tau\Phi(\tau,f)$, while already proven equality (\ref{razl}) reads as $(A(\tau)-I)f=\tau\partial f+\tau^2\Theta(\tau,f)$ and  (\ref{octhet}) says that $\|\Theta(\tau,f)\|\leq C(f)$. So by setting $\Phi(\tau,f)=\tau\Theta(\tau,f)$ base of induction ($k=1$) is proven because $\|\Phi(\tau,f)\|=|\tau|\cdot\|\Theta(\tau,f)\|\leq |\tau|C(f)\to0$ as $\tau\to 0$.

b) Suppose that for some $k-1\geq 1$ for each $g\in C_0^{ebd}(\mathbb{R})$ we have 
$$(A(\tau)-I)^{k-1}g=\tau^{k-1}\partial^{k-1}g+\tau^{k-1}\Phi(\tau,g),$$ where $\lim_{\tau\to0}\|\Phi(\tau,g)\|=0$. 

c) Let us derive (\ref{shagleft}) from b) and (\ref{razl}) for some $k\geq 2$ and fixed $f\in C_0^{ebd}(\mathbb{R})$. Firstly, let us introduce expressions $M_1$, $M_2$, $M_3$, $M_4$ by performing the following transformations:
$$
(A(\tau)-I)^kf=(A(\tau)-I)^{k-1}(A(\tau)-I)f\stackrel{(\ref{razl})}{=}
$$
$$
=(A(\tau)-I)^{k-1}\left(\tau\partial+\dots+(\tau\partial)^{k}/k!+\tau^{k+1}\Theta(\tau)\right)f
=\underbrace{\tau(A(\tau)-I)^{k-1}\partial f}_{=M_1}+$$
$$+\underbrace{\sum_{j=2}^{k-1}\frac{\tau^j}{j!}(A(\tau)-I)^{k-1}\partial^jf}_{=M_2}
+\underbrace{\frac{\tau^k}{k!}(A(\tau)-I)^{k-1}\partial^kf}_{=M_3}
+\underbrace{\tau^{k+1} (A(\tau)-I)^{k-1}\Theta(\tau)f}_{=M_4}.
$$
For $k=2$ the sum $M_2$ is empty and thus equals to zero. Below we will show that $M_1=\tau^k\partial^kf+o(\tau^k)$ in step c-1), that $M_2=o(\tau^k)$ in step c-2), that $M_3=o(\tau^k)$ in step c-3) and that $M_4=o(\tau^k)$ in step c-4). This will be enough to finish the induction process. Then in c-5) we count how many bounded derivatives we need function $f$ to have, specifying the smallest space $C_0^{ebd}(\mathbb{R})$. After that we formulate proven statement without the symbol $C_0^{ebd}(\mathbb{R})$.

c-1) As $f\in C_0^{ebd}(\mathbb{R})$ implies $\partial f\in C_0^{ebd}(\mathbb{R})$ and $\partial^k f\in C_0^{ebd}(\mathbb{R})$, then we can set $g=\partial f$ in b) and obtain
$$M_1=\tau(A(\tau)-I)^{k-1}\partial f\stackrel{b)}{=}\tau\tau^{k-1}\partial^{k-1}\partial f+\tau\tau^{k-1}\Phi(\tau,\partial f)=\tau^k\partial^kf+o(\tau^k)$$
because $f$ is fixed, and due to b) we have  $\lim_{\tau\to0}\|\Phi(\tau,\partial f)\|=0$.

c-2) In the same way as in step c-1), we mention that $f\in C_0^{ebd}(\mathbb{R})$ implies that $\partial^2 f\in C_0^{ebd}(\mathbb{R})$, $\dots$ , $\partial^{k-1} f\in C_0^{ebd}(\mathbb{R})$, and for $j=2,\dots,k-1$ we have $\partial^{j} f\in C_0^{ebd}(\mathbb{R})$ and  $\partial^{k+j-1} f\in C_0^{ebd}(\mathbb{R})$. So for $j=2,\dots,k-1$ we can ($k-2$ times, under the summation sign) set $g=\partial^jf$ in b) and obtain
$$
M_2=\sum_{j=2}^{k-1}\frac{\tau^j}{j!}(A(\tau)-I)^{k-1}\partial^jf=\sum_{j=2}^{k-1}\frac{\tau^j}{j!}\big(\tau^{k-1}\partial^{k-1}\partial^jf+\tau^{k-1}\Phi(\tau,\partial^jf)\big)=
$$
$$
=\tau^{k+1}\sum_{j=2}^{k-1}\frac{1}{j!}\tau^{j-2}\partial^{k+j-1}f + \tau^{k+1}\sum_{j=2}^{k-1}\frac{1}{j!}\tau^{j-2}\Phi(\tau,\partial^jf)=o(\tau^k)+o(\tau^k)=o(\tau^k)
$$
because $f$ is fixed and because due to b) we have  $\lim_{\tau\to0}\|\Phi(\tau,\partial^j f)\|=0$ for each $j=2,\dots,k-1$.

c-3) We need to estimate $M_3=\frac{1}{k!}\tau^k(A(\tau)-I)^{k-1}\partial^kf$. Recall that in the proof of (CT2) we have shown that $A(0)=I$, and in the proof of $(CT1)$ that the mapping $\tau\longmapsto A(\tau)h$ is continuous for each $h\in L_2(\mathbb{R})$, so $\lim_{\tau\to0}(A(\tau)-I)h=0$ for each $h\in L_2(\mathbb{R})$.

c-3-i) If $k=2$, then we set $h=\partial^2f$ and obtain $\lim_{\tau\to0}(A(\tau)-I)^{2-1}\partial^2f=0$, and $M_3=\frac{1}{2!}\tau^2(A(\tau)-I)^{2-1}\partial^2f=o(\tau^2)$.

c-3-ii) If $k> 2$, then recall the proof of item 1), where we have shown that $\|A(\tau)\|=1$, so $\|(A(\tau)-I)\|\leq \|A(\tau)\|+\|I\|=2$. Then we have $\|M_3\|=\frac{1}{k!}\|\tau^k(A(\tau)-I)^{k-1}\partial^kf\|=\frac{1}{k!}|\tau|^k\|(A(\tau)-I)^{k-2}(A(\tau)-I)\partial^kf\|\leq\frac{1}{k!} |\tau|^k\|(A(\tau)-I)\|^{k-2}\|(A(\tau)-I)\partial^kf\|\leq\frac{1}{k!} |\tau|^k2^{k-2}\|(A(\tau)-I)\partial^kf\|=o(\tau^k)$ because we can set $h=\partial^kf$ and obtain  $\lim_{\tau\to0}\|(A(\tau)-I)\partial^kf\|=0$ as before. 

c-4) The fact that $M_4=o(\tau^k)$ follows from the following estimation: $$\|M_4\|= \left\|\tau^{k+1} (A(\tau)-I)^{k-1}\Theta(\tau)f\right\|$$
$$\leq |\tau|^{k+1}\|(A(\tau)-I)\|^{k-1}\|\Theta(\tau)f\|\stackrel{(\ref{octhet})}{\leq}|\tau|^{k+1}2^{k-2}C(f),$$ 
which we get by using the inequality $\|(A(\tau)-I)\|\leq2$ proven in c-3-ii). So induction on  $k$ is finished.

c-5) The expression with highest derivative of $f$ is  $\partial^{k+j-1}f$ which appears in step c-2) for $j=k-1$. So function $\partial^{k+j-1}f=\partial^{k+k-1-1}f=\partial^{2k-2}f$ must have enough bounded derivatives. As we do not use derivatives of $\partial^{2k-2}f$ explicitly, we conclude that two bounded derivatives is enough for it. So $\partial^{2k-2}f\in C_0^2(\mathbb{R})$ hence $f\in C_0^{2+2k-2}(\mathbb{R})=C_0^{2k}(\mathbb{R})$. Now the statement of iv) is completely proven.

v). Recall that $(B_{a_k}\varphi)(x)=a_k(x)\varphi(x)$, where function $a_k$ is bounded together with its derivatives up to order $2k$, where $k$ ranges from $1$ to $K$, and $K$ is fixed in the very beginning of the paper. So if $\varphi\in C_0^\infty(\mathbb{R})$ then  $B_{a_k}\varphi\in C_0^{2k}(\mathbb{R})$, and we can set $B_{a_k}\varphi=f$ in (\ref{shagleft}) and obtain
\begin{equation}\label{pochti}
(A(\tau)-I)^kB_{a_k}\varphi=\tau^k\partial^kB_{a_k}\varphi+\tau^k\Phi_1(\tau,\varphi)
\end{equation}
where $\Phi_1(\tau,\varphi)=\Phi(\tau,B_{a_k}\varphi)\to 0$ as $\tau\to0$ due to iv).

vi). Considering the Taylor's formula and reasoning analogous to iii), one needs to substitute $\tau$ by $-\tau$, which allows to obtain the following representation:
\begin{equation}\label{razl-1}
(I-A(\tau)^*)\varphi = \left(\tau\partial-\dots-\frac{(-\tau\partial)^{m+k}}{(m+k)!}+\tau^{m+k+1}\Theta_2(\tau)\right)\varphi
\end{equation}
for each $\varphi\in C_0^\infty(\mathbb{R})$, $m=0,1,2,\dots$, and $k=1,2,\dots$, where operator $\Theta_2(\tau)\colon C_0^\infty(\mathbb{R})\to C_0^\infty(\mathbb{R})$ is linear and 
\begin{equation}\label{octhet-1}
\sup_{\tau\in\mathbb{R}}\|\Theta_2(\tau)\varphi\|\leq C_2(\varphi).
\end{equation}

vii). By induction on $m=0,1,2,\dots$ (as was done above with $k=1,2,\dots$) we prove that
\begin{equation}\label{pochti-m}
(A(\tau)-I)^kB_{a_k}(I-A(\tau)^*)^m\varphi=\tau^{k+m}\partial^kB_{a_k}\partial^m\varphi+\tau^{k+m}\Phi_3(\tau,\varphi),
\end{equation}
where $\Phi_3(\tau,\varphi)\to 0$ as $\tau\to0$. This time induction is simpler than it was in iv) as we need not care about the class of differentiability: $\varphi$ has bounded derivatives of all orders which we have already  used in vi). We acted carefully in iv) because function $a_k$ that arises in v) had bounded derivatives only up to order $2k$. 

a) For $m=0$ the desired equality (\ref{pochti-m}) coincides with already proven equality (\ref{pochti}).

b) Suppose that for some $m\geq1$ and all $\psi\in C_0^\infty(\mathbb{R})$ we have
\begin{equation}\label{pochti-m-predp}
(A(\tau)-I)^kB_{a_k}(I-A(\tau)^*)^{m-1}\psi=\tau^{k+m-1}\partial^kB_{a_k}\partial^{m-1}\psi+\tau^{k+m-1}\Phi_3(\tau,\psi)
\end{equation}
where $\Phi_3(\tau,\psi)\to 0$ as $\tau\to0$.

c) Let us derive (\ref{pochti-m}) for some $m\geq 1$ from (\ref{pochti-m-predp}) and (\ref{razl-1}). Indeed
$$
(A(\tau)-I)^kB_{a_k}(I-A(\tau)^*)^m\varphi=(A(\tau)-I)^kB_{a_k}(I-A(\tau)^*)^{m-1}(I-A(\tau)^*)\varphi$$
$$\stackrel{(\ref{razl-1})}{=}(A(\tau)-I)^kB_{a_k}(I-A(\tau)^*)^{m-1}\left(\tau\partial
-\sum_{j=2}^{m+k}\frac{(-\tau\partial)^{j}}{j!}+\tau^{m+k+1}\Theta_2(\tau)\right)\varphi$$
$$
=\underbrace{(A(\tau)-I)^kB_{a_k}(I-A(\tau)^*)^{m-1}\tau\partial\varphi}_{=P_1} -$$ $$-\underbrace{(A(\tau)-I)^kB_{a_k}(I-A(\tau)^*)^{m-1}\sum_{j=2}^{m+k}\frac{(-\tau\partial)^{j}}{j!}\varphi}_{=P_2}+ 
$$
$$+\underbrace{(A(\tau)-I)^kB_{a_k}(I-A(\tau)^*)^{m-1}\tau^{m+k+1}\Theta_2(\tau)\varphi}_{=P_3}.$$
We will show that $P_1=\tau^{k+m}\partial^kB_{a_k}\partial^m\varphi+o(\tau^{k+m})$ in step c-1), that $P_2=o(\tau^{k+m})$ in step c-2), and that $P_3=o(\tau^{k+m})$ in step c-3).

c-1). We can set $\psi=\partial\varphi$ in (\ref{pochti-m-predp}) and obtain   $P_1=(A(\tau)-I)^kB_{a_k}(I-A(\tau)^*)^{m-1}\tau\partial\varphi=\tau\tau^{k+m-1}\partial^kB_{a_k}\partial^{m-1}\partial\varphi+\tau\tau^{k+m-1}\Phi_3(\tau,\partial\varphi)=\tau^{k+m}\partial^kB_{a_k}\partial^m\varphi$ $+o(\tau^{k+m})$ because $\varphi\in C_0^\infty(\mathbb{R})$ is fixed and  $\lim_{\tau\to0}\Phi_3(\tau,\partial\varphi)=0$ due to b).

c-2). For $j=2,\dots,m+k$ we can ($m+k-1$ times, under the summation sign) set $\psi=\partial^j\varphi$ in b) and obtain

$P_2=(A(\tau)-I)^kB_{a_k}(I-A(\tau)^*)^{m-1}\sum_{j=2}^{m+k}\frac{(-\tau\partial)^{j}}{j!}\varphi=\tau^2\sum_{j=2}^{m+k}\frac{(-1)^{j}\tau^{j-2}}{j!}$\\
$(A(\tau)-I)^kB_{a_k}(I-A(\tau)^*)^{m-1}\partial^j\varphi\stackrel{(\ref{pochti-m-predp})}{=}
\tau^2\sum_{j=2}^{m+k}\Big(\frac{(-1)^{j}\tau^{j-2}}{j!}
\tau^{k+m-1} \partial^kB_{a_k}\partial^{m-1}\partial^j\varphi$ $+\tau^{k+m-1}\Phi_3(\tau,\partial^j\varphi)\Big)=\tau^{k+m+1}\sum_{j=2}^{m+k}\frac{(-1)^{j}\tau^{j-2}}{j!}
\partial^kB_{a_k}\partial^{m+j-1}\varphi+\tau^{k+m+1}
\sum_{j=2}^{m+k}$ $\Phi_3(\tau,\partial^j\varphi)=o(\tau^{k+m})+o(\tau^{k+m+1})=o(\tau^{k+m})$ because $\varphi\in C_0^\infty(\mathbb{R})$ is fixed and  $\lim_{\tau\to0}\Phi_3(\tau,\partial^j\varphi)=0$ due to b).

c-3). Recall that operator $B_{a_k}$ is bounded and $\|B_{a_k}\|\leq\sup_{z\in\mathbb{R}}|a_k(z)|$. Also $\|A(\tau)\|=\|A(\tau)^*\|=1$ so $\|A(\tau)-I\|\leq 2$ and $\|I-A(\tau)^*\|\leq 2$. Due to (\ref{octhet-1}) we have $\|\Theta_2(\tau)\varphi\|\leq C(\varphi)$ for all $\tau\in\mathbb{R}$. Using all that we can estimate
$\|P_3\|=\|(A(\tau)-I)^kB_{a_k}(I-A(\tau)^*)^{m-1}\tau^{m+k+1}\Theta_2(\tau)\varphi\|\leq \tau^{m+k+1}\|A(\tau)-I\|^k\|B_{a_k}\|\cdot \|I-A(\tau)^*\|^{m-1}\|\Theta_2(\tau)\varphi\|\leq \tau^{m+k+1} 2^k\sup_{z\in\mathbb{R}}|a_k(z)|\cdot2^{m-1}C(\varphi)=\tau^{m+k+1}\cdot\mathrm{const}=o(\tau^{m+k})$. Induction on $m$ if finished and (\ref{pochti-m}) is now proved.

viii). If we  set $m=k\geq 1$ and $\tau=t^{1/2k}$ in (\ref{pochti-m}) and recall the definition of $F_k$ in (\ref{FSdef}) we obtain
$F_k(t)\varphi=(A(t^{1/2k})-I)^kB_{a_k}(I-A(t^{1/2k})^*)^k\varphi=(t^{1/2k})^{k+k}\partial^kB_{a_k}\partial^k\varphi+(t^{1/2k})^{k+k}\Phi_3(t^{1/2k},\varphi)=t\partial^kB_{a_k}\partial^k\varphi+t\Phi_3(t^{1/2k},\varphi)=t\partial^kB_{a_k}\partial^k\varphi+o(t)$ because $\lim_{t\to0}\Phi_3(t^{1/2k},\varphi)=0$ as follows from (\ref{pochti-m}), theorem on change of variables in a limit, and the fact that $\lim_{t\to0}t^{1/2k}=0$. 

Recalling that operator $\partial$ is the differentiation operator $(\partial f)(x)=f'(x)$, and $B_{a_k}$ is multipication by $a_k$, we have proved that for each $k=1,\dots,K$
$$
(F_k(t)\varphi)(x)=t\frac{d^k}{dx^k}\left(a_k(x) \frac{d^k}{d x^k} \varphi(x)\right)+o(t).$$
Repeating what have been said in step i) we finish the proof of (CT3).

(CT4) is true due to the following condition of the theorem: the operator $\mathcal{H}$ defined on $C_0^\infty(\mathbb{R})$ is essentially self-adjoint in $L_2(\mathbb{R})$, i.e. operator $(\mathcal{H}, C_0^\infty(\mathbb{R}))$ is closable and its closure --- let us denote it as $(\mathcal{H},Dom(\mathcal{H}))$ --- is a self-adjoint operator. To prove (CT4) we set $\mathcal{D}=C_0^\infty(\mathbb{R})$ as in (CT3); note that here (in the proof of (CT4)) we don't use the fact that $(\mathcal{H},Dom(\mathcal{H}))$ is self-adjoint and don't describe the set $Dom(\mathcal{H})\subset L_2(\mathbb{R})$. Item 2) of the theorem is thus proved.

Item 3). We will prove that $S(t)^*=S(t)$ for all $t\geq 0$ using the following general fact:  $(\alpha+\beta)^*=\alpha^*+\beta^*$ and $(\alpha\beta)^*=\beta^*\alpha^*$ for all linear bounded operators $\alpha$ and $\beta$ in Hilbert space. Recall that due to the item 1) (which is now proven) all of the summands in  formula (\ref{FSdef}) which defines $S(t)=I+F_0(t)+F_1(t)+\dots+F_K(t)$ are bouned and defined on the whole space $L_2(\mathbb{R})$. So $S(t)^*$ exists and is also defined on the whole space $L_2(\mathbb{R})$. 

Operator $F_0(t)$ is self-adjoint because it is an operator of multiplication by a real-valued bounded function $x\longmapsto w(ta_0(x))$.

For $k\geq 1$ we can use the representation $F_k(t)=X(t)^kB_{a_k}Y(t)^k$, where $X(t)=A(t^{1/2k})-I$, $Y(t)=I-A(t^{1/2k})^*$ to see that $X(t)^*=-Y(t)$ and $Y(t)^*=-X(t)$. Note also that $B_{a_k}^*=B_{a_k}$ because $B_{a_k}$ is an operator of multiplication by a bounded real-valued function $a_k$. So $F_k(t)^*=(X(t)^kB_{a_k}Y(t)^k)^*=(B_{a_k}Y(t)^k)^*((X(t)^k)^*=(Y(t)^*)^kB_{a_k}^*(X(t)^*)^k=(-X(t))^k$ $B_{a_k}^*(-Y(t))^k=(-1)^{k+k}X(t)^kB_{a_k}Y(t)^k=F_k(t)$.  
Then $S(t)^*=(I+F_0(t)+F_1(t)+\dots+F_K(t))^*=I^*+F_0(t)^*+F_1(t)^*+\dots+F_K(t)^*=I+F_0(t)+F_1(t)+\dots+F_K(t)=S(t)$ and item 3) is proved.

Item 4). We have proved in item 1) that  $\|-iF(t)\|=\|F(t)\|<\infty$ for each $t\geq 0$. Then $\sum_{j=0}^\infty\|-iF(t)\|^j/j!<\infty$ implies that the power series  $\sum_{j=0}^\infty (-iF(t))^j/j!$  converges in $\mathscr{L}(\mathcal{F})$, so linear bounded operator $\exp(-iF(t))=\sum_{j=0}^\infty(-iF(t))^j/j!\in\mathscr{L}(\mathcal{F})$ is well-defined, see details in \cite{EN1}.

Item 5) follows directly from Stone's theorem (theorem \ref{Stth}) and the fact operator $(\mathcal{H},Dom(\mathcal{H}))$ is self-adjoint. See also proof of (CT4).

Item 6) follows from theorem \ref{RemQ} and items 2) and 3) which are proven above.

Item 7) follows from item 6) and the general fact of $C_0$-semigroup theory \cite{EN1}: if operator $L$ is a generator of $C_0$-semigroup $\left(e^{tL}\right)_{t\geq 0}$ of linear bounded operators in Banach space, then the Cauchy problem $u'(t)=Lu(t), u(0)=u_0$ for Banach-space-valued function $u$ has a unique solution $u(t)=e^{tL}u_0$ for each $u_0$ from this Banach space. In the present theorem the role of the Banach space is played by $L_2(\mathbb{R})$, $u_0$ is $\psi_0$ and $u(t)=\psi(t,\cdot)$ i.e. $(u(t))(x)=\psi(t,x)$. In $L_2(\mathbb{R})$ two functions $f$ and $g$ are equal iff $f(x)=g(x)$ for almost all $x\in\mathbb{R}$. $\Box$ 

\begin{remark}
The following formulas may be useful in computation of the right-hand side of the solution-giving formula from item 7) of the above theorem:
$$F_k(t)=\left(A(t^{1/2k})-I\right)^kB_{a_k}\left(I-A(t^{1/2k})^*\right)^k=$$
$$=\sum_{j_2=0}^k\frac{k!(-1)^{k-j_2}}{j_2!(k-j_2)!}A(t^{1/2k})^{j_2}B_{a_k}\sum_{j_1=0}^k\frac{k!(-1)^{j_1}}{j_1!(k-j_1)!}A(t^{1/2k})^{*j_1}=$$
$$=\sum_{j_1=0}^k\frac{k!(-1)^{j_1}}{j_1!(k-j_1)!}\sum_{j_2=0}^k\frac{k!(-1)^{k-j_2}}{j_2!(k-j_2)!}A(t^{1/2k})^{j_2}B_{a_k}A(t^{1/2k})^{*j_1},$$
where
$$
\left(A(t^{1/2k})^{j_2}B_{a_k}A(t^{1/2k})^{*j_1}f\right)(x)=a_k\left(x+j_2t^{1/2k}\right)f\left(x+(j_2-j_1)t^{1/2k}\right).
$$
These formulas are derived directly from the Newton's binomial formula and definitions of the operators used, see formula (\ref{FSdef}) and ones just above it.
\end{remark}

\begin{remark}\label{empty-rem} Note that the solution of (\ref{CP1}) is provided only with the following assumption for coefficients $a_k$, $k=0,1,\dots,K$: the operator $\mathcal{H}$ defined on $C_0^\infty(\mathbb{R})$ is essentially self-adjoint in $L_2(\mathbb{R})$, i.e. the operator $(\mathcal{H}, C_0^\infty(\mathbb{R}))$ is closable and its closure --- let us denote it as $(\mathcal{H},Dom(\mathcal{H}))$ --- is self-adjoint.  One might ask themselves if the set of coefficients satisfying this condition may be empty. Theorem \ref{lm-RS} shows (one needs to set $d=1$) that at least for $K=1$ and the operator $\mathcal{H}$ of the form $(\mathcal{H}f)(x)=-f''(x)+v(x)f(x)$ this condition is satisfied for  $a_1(x)\equiv -1$ and $a_0(x)=v(x)$.   
\end{remark}

\begin{theorem}\label{lm-RS}(theorem  X.28 in \cite{RS2}) Let $v\in L^{loc}_2(\mathbb{R}^d)$ with $v\geq 0$ pointwise. Then $-\Delta +v$ is essentially self-adjoint on $L_2(\mathbb{R}^d)$.
\end{theorem}

Here "$v\geq 0$ pointwise" means that $v(x)\geq 0$ for each $x\in \mathbb{R}^d$ (see the discussion in the beginning of the chapter X.4 of \cite{RS2}).

\subsection{Example: Schr\"odinger equation with  Sturm-Liouville operator}

Let us consider the particular case of $K=1$. We recall some notation to make this subsection independent of other text of the present paper, but recommend to see remark \ref{empty-rem} above. Then, for $K = 1$ $(\mathcal{H}\varphi)(x)=q(x)\varphi(x)+(p(x)\varphi'(x))'$ where we denoted $a_0=q$, $a_1=p$ just as in section of heuristic arguments (section \ref{heur-arg}). Function $p$ is twice differentiable and bounded together with its derivatives. Measurable function $q$ may be unbounded but its square is locally integrable. Then the Cauchy problem that we are solving is 
\begin{equation}\label{ShLex}
\left\{ \begin{array}{ll}
i\psi'_t(t,x)=(p(x)\psi'_x(t,x))'_x+q(x)\psi(t,x)=\mathcal{H}\psi(t,x),\\ 
\psi(0,x)=\psi_0(x).
\end{array} \right.
\end{equation}

As before, to find the solution we use the following families (parametrized by $t\geq 0$) of linear bounded operators in $L_2(\mathbb{R})$:
$$(B_pf)(x)=p(x)f(x), (A(t)f)(x)=f(x+t), (A(t)^*f)(x)=f(x+t),
$$
$$
(F_0(t)f)(x)=w(tq(x))f(x), F_1(t)=(A(\sqrt{t})-I)B_p(I-A(\sqrt{t})^*),
$$
$$
F(t)=F_0(t)+F_1(t)=F_0(t)+A(\sqrt{t})B_p-A(\sqrt{t})B_pA(\sqrt{t})^*-B_p+B_pA(\sqrt{t})^*.
$$
So 
\begin{multline}\label{sem-sdv}
(F(t)f)(x)=w(tq(x))f(x)+p(x+\sqrt{t})f(x+\sqrt{t})\\
-p(x+\sqrt{t})f(x)-p(x)f(x)+p(x)f(x-\sqrt{t}).
\end{multline}

Theorem \ref{mainth} then says that for each $\psi_0\in L_2(\mathbb{R})$ solution of (\ref{ShLex}) exists for all $t\geq 0$ and for almost all $x\in\mathbb{R}$ satisfies the formula 
\begin{equation}\label{ShLsol}
\psi(t,x)=\left(e^{-it\mathcal{H}}\psi_0 \right)(x)=\left(\lim_{n\to\infty}\lim_{j\to+\infty}\sum_{k=0}^{j}\frac{(-in)^k}{k!}F(t/n)^k\psi_0\right)(x),
\end{equation}
where $F(t/n)$ is obtained by substitution of $t$ with $t/n$ in (\ref{sem-sdv}), and $F(t/n)^k$ is a compisition of $k$ copies of the linear bounded operator $F(t/n)$.

\subsection{Quasi-Feynman formulas with delta-functions}\label{quasf-1}

Formal equlity $f(h)=\int_{\mathbb{R}}\delta(y-h)f(y)dy$ with Dirac's delta-function holds for each $h\in\mathbb{R}$. Using it we obtain the formula 
$$f(x+\sqrt{t})=\int_{\mathbb{R}}\delta(y-x-\sqrt{t})f(y)dy,$$ and after a change of variables  $y=x+z,  z=y-x, dy=dz$ we have 
$$f(x+\sqrt{t})=\int_{\mathbb{R}}\delta(z-\sqrt{t})f(x+z)dy,$$ 
which, after setting $t=0$, gives us
$$f(x)=\int_{\mathbb{R}}\delta(z)f(x+z)dy.$$ 
Using this approach one can rewrite the equality (\ref{sem-sdv}) in the following form:
$$
(F(t)f)(x)=\int_\mathbb{R}\Phi(z,x,t)f(x+z)dz,
$$
where
\begin{multline}\label{Pfform}
\Phi(z,x,t)=\left(w(tq(x))-p(x+\sqrt{t})-p(x)\right)\delta(z)\\
+p(x+\sqrt{t})\delta(z-\sqrt{t})+p(x)\delta(z+\sqrt{t}).
\end{multline}
Let us calculate $F(t/n)^k$ using this representation. Let's say $q=2$:
\begin{multline*}
(F(t/n)^2f)(x)=\int_\mathbb{R}\Phi(z_1,x,t/n)(F(t/n)f)(x+z_1)dz_1\\
=\int_\mathbb{R}\Phi(z_1,x,t/n)
\int_\mathbb{R}\Phi(z_2,x+z_1,t/n)f(x+z_1+z_2)dz_2dz_1.
\end{multline*}

Finally, (\ref{ShLsol}) reads as
\begin{multline}\label{quasi-f-ShL}
\psi(t,x)=
\lim\limits_{n\to\infty}
\lim\limits_{j\to+\infty}\sum_{k=0}^{j}\frac{(-in)^k}{k!}
\int_{\mathbb{R}}\Phi(z_1,x,t/n)\\ \int_{\mathbb{R}}\Phi(z_2,x+z_1, t/n)\int_{\mathbb{R}}\Phi(z_3,x+z_1+z_2, t/n)\dots\\ 
\dots\int_{\mathbb{R}}\Phi(z_k,x+z_1+\dots+z_{k-1}, t/n)\psi_0(x+z_1+\dots+z_k)dz_k\dots dz_1.
\end{multline}

On the right-hand side of (\ref{quasi-f-ShL}) we have a sum of multiple integrals with multiplicity tending to infinity. Such expressions appeared first in \cite{R1} and were named quasi-Feynman formulas. Quasi-Feynman formulas in (\ref{quasi-f-ShL}) have distributions under the integral sign, as expression (\ref{Pfform}) for $\Phi$ includes Dirac's delta-function, see \cite{R7}. 

Note that the right-hand side of (\ref{quasi-f-ShL}) is a formal expression that is well-defined thanks to formula (\ref{ShLsol}) and the equality $f(h)=\int_{\mathbb{R}}\delta(y-h)f(y)dy$. However, we can forget about formula (\ref{ShLsol}) and study (\ref{quasi-f-ShL}) independently using distribution theory. For example, one might be interested in finding out the following: 1) do the limits in (\ref{quasi-f-ShL}) exist in the topology on the space of distributions? 2) can one go from an iterated integral to a multiple intgeral? 3) can one change the order of integration?

\subsection{Multi-dimensional Schr\"odinger equation}

In this section we fix an arbitrary  $d\in\mathbb{N}$, use an arbitrary non-zero real number $a$ as a parameter (one can set $a=1$ or $a=-1$ depending on the preferable way of writing the Schr\"odinger equation), and study the Cauchy problem in the space $\mathcal{F}=L_2(\mathbb{R}^d)$ over the field $\mathbb{C}$ as follows:
\begin{equation}\label{f_Koshi-md}
\begin{cases}
i\psi'_{t}(t,x)=-\frac{1}{2}a\sum\limits_{m=1}^d\psi_{x_mx_m}''(t,x)+aV(x)\psi(t,x), & t\in\mathbb{R}^1, x\in\mathbb{R}^d, \\
\psi(0,x)=\psi_{0}(x), & x\in\mathbb{R}^d.
\end{cases}
\end{equation} 

We prove that a solution of this Cauchy problem exists under some assumptions and provide a formula that expresses this solution in terms of parameter $a$, initial condition $\psi_0$ and the potential $V$. Our method is general enough to cover most of the cases that are useful for physics. In particular, one can set $V$ to be any non-negative continuous function, including potential a of quantum harmonic oscillator ($V(x)=\|x\|^2$) and potentials of the two most known  quantum anharmonic oscillators ($V(x)=\|x\|^4$, $V(x)=\|x\|^2+\|x\|^4$). Our technique is also applicable to degenerate ($V(x)=0$ for some $x\neq 0$) and non-smooth potentials. The main result of this subsection is the following theorem.

\begin{theorem}\label{thRd} (Announced in short communication \cite{R6} without full proof). Suppose that  function  $V\colon\mathbb{R}^d\to\mathbb{R}$ belongs to the space $L^{loc}_2(\mathbb{R}^d)$, i.e. $V$ is measurable and  $\int_{\|x\|\leq R}V(x)^2dx<\infty$ for each $R>0$, where $\|x\|=(x_1^2+\dots+x_d^2)^{1/2}$. Suppose that $a\in\mathbb{R}, a\neq 0$. Suppose that function $w\colon\mathbb{R}\to\mathbb{R}$ is bounded (denote  $M=\sup_{x\in\mathbb{R}}|w(x)|$), continuous,  differentiable at zero and  $w(0)=0$, $w'(0)=1$; for example, one can take  $w(x)=\sin(x)$,  $w(x)=\arctan(x)$, $w(x)=\tanh(x)=(e^x-e^{-x})/(e^x+e^{-x})$ etc. Suppose that for each $j=1,\dots,d$ constant vector $e_j\in\mathbb{R}^d$ has $1$ at position $j$ and has $0$ at other $d-1$ positions. For each function  $f\in L_2(\mathbb{R}^d)$, each smooth function  $\varphi\colon\mathbb{R}^d\to\mathbb{C}$ and each $x\in\mathbb{R}$, $t\geq 0$ define
\begin{multline}\label{oper-sdvig}(W(t)f)(x)=\frac{1}{2d}\sum_{j=1}^d\Big[f\left(x+\sqrt{d}\sqrt{t}e_j\right)\\
+f\left(x-\sqrt{d}\sqrt{t}e_j\right)-2f(x)\Big]+w(-tV(x))f(x),
\end{multline}
\begin{equation}\label{oper-H}(H\varphi)(x)=\frac{1}{2}\Delta\varphi''(x)-V(x)\varphi(x).\end{equation}
Suppose also that at least one of these two conditions is satisfied: A) if we use the symbol $C_0^\infty(\mathbb{R}^d)$ for the set of all infinitely smooth functions $\mathbb{R}^d\to\mathbb{R}$ with compact support then the closure of the operator  $(H,C_0^\infty(\mathbb{R}^d))$ is a self-adjoint operator in  $L_2(\mathbb{R}^d)$; B) $V(x)\geq 0$ for all $x\in\mathbb{R}^d$.

Then:  

1) For each $t\geq 0$ we have $\|W(t)\|\leq 2+M$. 

2) For each $t\geq 0$ we have $W(t)=W(t)^*$. 

3) $G(t)=W(t)+I$ is Chernoff-tabgent to $H$, where $I$ is the identity operator in $L_2(\mathbb{R}^d)$.

4) There exists a $C_0$-group $(e^{iatH})_{t\in\mathbb{R}}$ of linear bounded unitary operators in $L_2(\mathbb{R}^d)$.

5) Cauchy problem (\ref{f_Koshi-md}) for Schr\"odinger equation can be rewritten as
\begin{equation}\label{f_Koshi-md-new}
\begin{cases}
\psi'_{t}(t,x)=iaH\psi(t,x), & t\in\mathbb{R}^1, x\in\mathbb{R}^d, \\
\psi(0,x)=\psi_{0}(x), & x\in\mathbb{R}^d.
\end{cases}
\end{equation} 
where the Hamiltonian is equal to $-aH$. For each $t\geq 0$ and $\psi_0\in L_2(\mathbb{R}^d)$ Cauchy problems (\ref{f_Koshi-md}) and (\ref{f_Koshi-md-new}) have a unique (in  $L_2(\mathbb{R}^d)$) solution $\psi(t,x)=\left(e^{iatH}\psi_0\right)(x)$, that continuously, with  respect to norm in $L_2(\mathbb{R}^d)$, depends (for fixed $t$) on $\psi_0$. For almost all $x\in\mathbb{R}^d$ and all $t\geq 0$ this solution satisfies the formula
\begin{equation}\label{fin_r2-md}
\psi(t,x)=\left(\lim_{n\to+\infty}\lim_{j\to+\infty}\sum_{k=0}^{j}\frac{(ian)^k}{k!}W(t/n)^k\psi_0\right)(x),
\end{equation}
where $W(t/n)$ is obtained by substitution of $t$ with $t/n$ in (\ref{oper-sdvig}), and $W(t/n)^k$ is a composition of $k$ copies of linear bounded operator $W(t/n)$.
\end{theorem}

\textbf{Proof.} 

Let us first show that condition A) follows from condition B) which is simpler to check. Indeed, as $V(x)\geq 0$ the function $v(x)=2V(x)$ satisfies conditions of theorem \ref{lm-RS}, so the operator $-\Delta+v$ on domain $C_0^\infty(\mathbb{R}^d)$ is essentially self-adjoint in $L_2(\mathbb{R}^d)$, i.e. it is closable and its closure is a self-adjoint operator. Hence the operator $H=-\frac{1}{2}(-\Delta+v)$ is also closable and its closure is a self-adjoint operator in $L_2(\mathbb{R}^d)$. Now let us prove theorem \ref{thRd}.

Item 1). Let us denote $$(A_j(t)f)(x)=f\left(x+\sqrt{d}\sqrt{t}e_j\right),\quad (B_j(t)f)(x)=f\left(x-\sqrt{d}\sqrt{t}e_j\right),$$ 
$$(C(t)f)(x)=w(-tV(x))f(x), \quad (If)(x)=f(x).$$ For arbitrary  $f\in L_2(\mathbb{R}^d)$ we can make a change of variable $y=x+\sqrt{d}\sqrt{t}e_j, dy=dx$ in the integral  $\|A_j(t)f\|=\left(\int_{\mathbb{R}^d}\left|f\left(x+\sqrt{d}\sqrt{t}e_j\right)\right|^2dx\right)^{1/2}=\left(\int_{\mathbb{R}^d}\left|f\left(y\right)\right|^2dy\right)^{1/2}=\|f\|$ so $\|A_j(t)\|=1$. The same reasoning shows that $\|B_j(t)\|=1$. For each  $f\in L_2(\mathbb{R}^d)$ we may estimate the integral as follows:  $\|C(t)f\|= \left(\int_{\mathbb{R}^d}\left|w(-tV(x))f(x)\right|^2dx\right)^{1/2}\leq \sup_{z\in\mathbb{R}}|w(z)| \left(\int_{\mathbb{R}^d}\left|f(x)\right|^2dx\right)^{1/2}=M\|f\|$ hence  $\|C(t)\|\leq M$. 

With the above notation we have  
\begin{equation}\label{Wt} W(t)=\frac{1}{2d}\sum_{j=1}^d\big(A_j(t)-2I+B_j(t)\big) +C(t),\end{equation}
hence for each $t\geq 0$ the following estimate is true:
$\|W(t)\|\leq\frac{1}{2d}\sum_{j=1}^d(\|A_j(t)\|$ $+2\|I\|+\|B_j(t)\|)+ \|C(t)\|\leq \frac{1}{2d}d (1+2+1) +M = 2+M$.

Item 2) follows from the fact that all the operators on the right-hand side of (\ref{Wt}) are bounded and $A_j(t)^*=B_j(t)$, $B_j(t)^*=A_j(t)$, $I^*=I$, $C(t)^*=C(t)$. 

Item 3). We set $\mathcal{F}=L_2(\mathbb{R}^d)$, $G(t)=W(t)+I$, $L=H$, $\mathcal{D}=C_0^\infty(\mathbb{R}^d)$ in definition (\ref{CTdef}) and check if for  $G$ and $H$ all the conditions of Chernoff tangency hold. Condition  (CT0) follows from assumptions of the theorem and item 1) that is proved above. Item 1) also gives us  $\|G(t)\|=\|W(t)+I\|\leq 3+M$, so the mapping $G\colon [0,+\infty)\to \mathscr{L}(L_2(\mathbb{R}))$ is well-defined.

(CT1). Let us fix $f\in L_2(\mathbb{R}^d)$ and prove that mappings $t\longmapsto A_j(t)f$, $t\longmapsto B_j(t)f$ and  $t\longmapsto C(t)f$ are continuous for each $j=1,\dots,d$.

(i). Let us first prove that $t\longmapsto A_j(t)f$ is continuous. Suppose that $t\geq 0, t_n\geq 0, t_n\to t$ and prove that $\|A_j(t)f- A_j(t_n)f\|\to0$. Note that we can make a change of variables $y=x+\sqrt{d}\sqrt{t}e_j, dy=dx$ in the integral:

$\|A_j(t)f- A_j(t_n)f\|^2=\int_{\mathbb{R}^d}\left|f\left(x+\sqrt{d}\sqrt{t}e_j\right)-f\left(x+\sqrt{d}\sqrt{t_n}e_j\right)\right|^2dx=\int_{\mathbb{R}^d}\left|f(y)-f\left(y-\sqrt{d}\sqrt{t}e_j+\sqrt{d}\sqrt{t_n}e_j\right)\right|^2dy,$ so we will not lose generality by assuming that $t=0$; indeed, conditions $\sqrt{t_n}-\sqrt{t}\to0$ and $\sqrt{t_n}\to0$ play the same role in the proof that follows. Setting $t=0$ in the definition of $A_j$ shows that $A_j(0)f=f$. 

Suppose that we are given an arbitrary $\varepsilon>0$. Let us find such $N\in\mathbb{N}$ that for all $n>N$ we have $\|f- A_j(t_n)f\|<\varepsilon$. The set $C_0^\infty(\mathbb{R}^d)$  of all infinitely smooth functions $\mathbb{R}^d\to\mathbb{R}$ with compact support is dense in  $L_2(\mathbb{R})$, so there exists such a function $g\in C_0^\infty(\mathbb{R}^d)$ that $\|f-g\|<\varepsilon/3.$ Then 
$\|f- A_j(t_n)f\|\leq \|f-g\| + \|g-A_j(t_n)g\| + \|A_j(t_n)(g-f)\|<\varepsilon/3 +\|g-A_j(t_n)g\| + 1\cdot \varepsilon/3$ because $\|A_j\|= 1$. Let us now find such $N$ that $\|g-A_j(t_n)g\|\leq\varepsilon/3$ holds for all $n>N$.

The fact that $g$ has compact support means that $g$ is zero outside some ball $B_R=\{x\in\mathbb{R}^d\colon \|x\|\leq R\}$ of radius $R$. As $g$ is uniformly continuous there exists such $\delta>0$, that $\|y-z\|<\delta$ implies   $|g(y)-g(z)|<\varepsilon/ \left(3\sqrt{vol(B_R)}\right).$ As $t_n\to 0$, there exists such  $N$ that for all $n>N$ we have $\sqrt{d}\sqrt{t_n}<\delta$. Then for all $n>N$ the following estimation is true:
$\|g-A(t_n)g\|=\left(\int_{\mathbb{R}^d} |g(x)- g(x+\sqrt{d}\sqrt{t_n}e_j)|^2dx \right)^{1/2}\leq \left(\int_{B_R} \left(\frac{\varepsilon}{3\sqrt{vol(B_R)}}\right)^2 dx\right)^{1/2} =\varepsilon/3.$ 
We have proved that function  $t\longmapsto A_j(t)f$ is continuous. 

(ii). By reasoning analogous  to (i) we can show that the mapping $t\longmapsto B_j(t)f$ is also continuous.

(iii). Now let us prove the continuity of the mapping $t\longmapsto C(t)f$. To achieve that we suppose that $t\geq 0, t_n\geq 0$, $\lim_{n\to\infty}t_n=t$ and prove that  $\lim_{n\to\infty}\|C(t)f-C(t_n)f\|=0$. From the definition of operator $C(t)$ we have the following:
\begin{equation}\label{integrand}
\|C(t_n)f-C(t)f\|^2=\int_{\mathbb{R}^d}|w(-t_nV(x))-w(-tV(x))|^2|f(x)|^2dx.\end{equation}
Then, $\sup_{z\in\mathbb{R}^d}|w(z)|= M$ imples that 
$\sup_{x\in\mathbb{R}^d} |w(-t_nV(x))-w(-tV(x))|^2\leq 4M^2<+\infty.$ From the continuity of $w$ we get that for each fixed  $x\in\mathbb{R}^d$ we have  $\lim_{n\to\infty}w(-t_nV(x))=w(-tV(x))$. So the sequence of integrands in (\ref{integrand}) is bounded by an integrable function $x\longmapsto 4M^2|f(x)|^2$ and converges to zero pointwise as  $n\to\infty$. 
Hence by Lebesgue dominated convergence theorem we have    $\lim_{n\to\infty}\int_{\mathbb{R}^d}|w(-t_nV(x))-w(-tV(x))|^2|f(x)|^2dx=0,$ which implies  continuity of the function $t\longmapsto C(t)f$. 

(iv). Recalling (\ref{Wt}) and steps (i), (ii), (iii) we can see that the function $t\longmapsto G(t)f$ is a finite linear combination of continuous functions $A_j$, $B_j$, $C$, so $t\longmapsto G(t)f$ is continuous and condition (CT1) is proven.

(CT2) is proven by setting $t=0$ in (\ref{oper-sdvig}), which implies $G(0)=W(0)+I=0+I=I$.

(CT3) we will show for a fixed arbitrary chosen function $f\in \mathcal{D}=C_0^\infty(\mathbb{R}^d)$; suppose that $R>0$ is such a number that  $f(x)=0$ for each $x\notin B_R=\{x\in\mathbb{R}^d: \|x\|\leq R\}$. 

(i) Let us first look at the first summand in   (\ref{oper-sdvig}). Due to the Taylor's expansion formula with remainders represented in Lagrange's form we have the following equalities for fixed $x\in\mathbb{R}^d$ and $t$ tending to zero (do not forget that $d\in\mathbb{N}$ so $dt$ is not a differential of $t$):
$$f(x+\sqrt{d}\sqrt{t}e_j)=f(x)+\sqrt{d}\sqrt{t}f_{x_j}'(x)+\frac{1}{2}dtf_{x_jx_j}''(x)+r_j^+(t,x),$$
$$f(x-\sqrt{d}\sqrt{t}e_j)=f(x)-\sqrt{d}\sqrt{t}f_{x_j}'(x)+\frac{1}{2}dtf_{x_jx_j}''(x)+r_j^-(t,x),$$
$$f(x+\sqrt{d}\sqrt{t}e_j)-2f(x)+f(x-\sqrt{d}\sqrt{t}e_j)=dtf_{x_jx_j}''(x)+r_j^+(t,x)+r_j^-(t,x).$$

Keeping in mind that function $f$ is zero outside the ball $B_R$ and has bounded third derivative, we come the following estimation:

$\|r_j^+(t,\cdot)+r_j^-(t,\cdot)\|=\left(\int_{\mathbb{R}^d}\left|r_j^+(t,x)+r_j^-(t,x)\right|^2dx \right)^{1/2}$ 

$\leq 2\Big(vol(B_R)\big(\frac{(td)^{3/2}}{3!}\sup_{z\in\mathbb{R}^d}\|f'''(z)\|\big)^2 \Big)^{1/2}=t\sqrt{t}\cdot\mathrm{const}=o(t)$. 

Summing by $j$, we obtain 
\begin{equation}\label{oci}\frac{1}{2d}\sum_{j=1}^d\left[f\left(x+\sqrt{d}\sqrt{t}e_j\right)+f\left(x-\sqrt{d}\sqrt{t}e_j\right)-2f(x)\right]=\frac{1}{2}t\Delta f(x)+o(t).
\end{equation}

(ii) Now let us examine the second summand in  (\ref{oper-sdvig}). Recall that function $w$ is bounded, continuous,  differentiable at zero and satisfies $w(0)=0$ and $w'(0)=1$. So, by Taylor's expansion formula with the remainder in Peano's form, $w$ can be represented as \begin{equation}\label{whreprez}w(z)=z+zh(z),\end{equation} 
where $\lim_{z\to0}h(z)=0$. Let us show that function $h$ is continuous and bounded. Let us define $h(0)=0$ and $h(z)=(w(z)-z)/z$ for $z\neq 0$. Function $w$ is continuous for all $z\in\mathbb{R}$, so $h$ is continuous for $z\neq 0$ due to the formula $h(z)=(w(z)-z)/z$, and $h$ is continuous at zero due to condition $\lim_{z\to0}h(z)=0=h(0)$. Now let us prove that $h$ is bounded.  Indeed, from $\lim_{z\to0}h(z)=0$ we get that $\sup_{|z|\leq 1}|h(z)|<\infty$. And for $|z|>1$ we can estimate  $|h(z)|=|w(z)/z - 1|\leq|w(z)/z| +1\leq |w(z)|+1<\infty$ because $w$ is bounded. So (\ref{whreprez}) and definition of $C(t)$ imply that  that for each $x\in\mathbb{R}^d$ and $z=-tV(x)$ we have
$$
(C(t))f(x)=w(-tV(x))f(x)\stackrel{(\ref{whreprez})}{=}-tV(x)f(x)-tV(x)f(x)h(-tV(x)).
$$
 
Now let us show that  $V(x)f(x)h(-t_nV(x))\to0$ in $L_2(\mathbb{R}^d)$, where $t_n\to 0$. Indeed, functions $f$ and $h$ are bounded, and $V\in L^{loc}_2(\mathbb{R}^d)$, so functions $x\longmapsto|V(x)f(x)h(-t_nV(x))|^2$ are: a) integrable on the ball $B_R$ (outside this ball $f$ is zero); b)  majorated on this ball by an integrable function $x\longmapsto|V(x)f(x)\sup_{z\in\mathbb{R}}|h(z)||^2$; c) converge to zero for each $x\in B_R$ as $n\to\infty$ because $\lim\limits_{z\to0}h(z)=0$. 

Then $\|V(\cdot)f(\cdot)h(-t_nV(\cdot))\|^2=\int_{\mathbb{R}^d}|V(x)f(x)h(-t_nV(x))|^2dx$\\$=\int_{B_R}|V(x)f(x)h(-t_nV(x))|^2dx\to0$ thanks to Lebesgue's dominated convergence theorem. So we have proved that  
\begin{equation}\label{ocii-1}
w(-tV(x))f(x)=-tV(x)f(x)+o(t).
\end{equation}

(iii) Summing (\ref{oci}) and (\ref{ocii-1}) and keeping in mind  (\ref{oper-sdvig}) and  (\ref{oper-H}), we arrive to the formula $W(t)f=tHf+o(t)$. (CT3) is now proved. 

(CT4) follows from condition A). The domain of the closure of the operator  $(H,C_0^\infty(\mathbb{R}^d))$ plays the role of $Dom(H)$ in (CT4). Item 3) of the theorem is now proved. 

Item 4) follows from the fact that (thanks to already proven items 2) and 3)) we can apply theorem \ref{RemQ} to $W$ and $H$.

Item 5) follows from theorem \ref{RemQ} and from the general fact of $C_0$-semigroup theory \cite{EN1}, which states that the Cauchy problem for a linear evolution equation  $u'(t)=Lu(t), u(0)=u_0$ has a unique in $\mathcal{F}$ solution $u(t)=e^{tL}u_0$ for each $u_0\in\mathcal{F}$, and this solution (for fixed $t$) depends on $u_0$ continuosly with respect to the norm in $\mathcal{F}$. Here we assume $\mathcal{F}=L^2(\mathbb{R}^d)$,  $L=iH$, $u_0=\psi_0$ and  $u(t)=\psi(t)$. 
$\Box$

\begin{remark} Let us now do the same for the $d$-dimensional case as what we have already done in section \ref{quasf-1} for the one-dimensional equation. Formal equality $f(h)=\int_{\mathbb{R}^d}\delta(y-h)f(y)dy$ with Dirac's delta-function holds for each $h\in\mathbb{R}^d$, which gives us
$$f(x+\sqrt{d}\sqrt{t}e_j)=\int_{\mathbb{R}^d}\delta(y-x-\sqrt{d}\sqrt{t}e_j)f(y)dy,$$ and after a change of variables  $y=x+z,  z=y-x, dy=dz$ we have 
$$f(x+\sqrt{d}\sqrt{t}e_j)=\int_{\mathbb{R}^d}\delta(z-\sqrt{d}\sqrt{t}e_j)f(x+z)dy.$$ Using this approach one can rewrite the equality (\ref{oper-sdvig}) in terms of distributions, and (\ref{fin_r2-md}) then reads as
\begin{multline}\label{distr-rav}\psi(t,x)=
\lim\limits_{n\to\infty}
\lim\limits_{j\to+\infty}\sum_{k=0}^{j}\frac{(ian)^k}{k!}
\int_{\mathbb{R}^d}\Phi(z_1,x,t/n)\\ \int_{\mathbb{R}^d}\Phi(z_2,x+z_1, t/n)\int_{\mathbb{R}^d}\Phi(z_3,x+z_1+z_2, t/n)\dots\\ 
\dots\int_{\mathbb{R}^d}\Phi(z_k,x+z_1+\dots+z_{k-1}, t/n)\psi_0(x+z_1+\dots+z_k)dz_k\dots dz_1,
\end{multline}
where
$$\Phi(z,x,t)=\frac{1}{2d}\sum_{j=1}^d[\delta(z-\sqrt{d}\sqrt{t}e_j) -2\delta(z)+ \delta(z+\sqrt{d}\sqrt{t}e_j)]+w(-tV(x))\delta(z).$$ 

A representation of the function $\psi$ in this form is a quasi-Feynman formula with generalized functions (=distributiuons) under the integral sign \cite{R7}. See also discussion in the end of section \ref{quasf-1}.
\end{remark}

\section*{Acknowledgements}

Author thanks his scientific advisor O.G.Smolyanov for profound consultations and words of encouragement, S.Mazzucchi for references, A.M.Nikiforov and D.V.Turaev for helpful discussions,  V.Zh.Sakbaev, N.N.Shamarov for comments. This work has been supported by the Basic Research Program at the HSE in 2018.


\begin{center}
	\textbf{References}
\end{center}

\begin{thebibliography}{0}

\bibitem{JL} Johnson G., Lapidus M. The Feynman Integral and Feynman Operational Calculus. ---
Clarendon Press, Oxford, 2000.

\bibitem{M1} Mazzucchi, S. Infinite Dimensional Oscillatory Integrals with Polynomial Phase and Applications to Higher-Order Heat-Type Equations --- Potential Analysis 49:2, pp 209-223, (2018).

\bibitem{M2} S. Bonaccorsi and S. Mazzucchi. High Order Heat-type Equations and Random Walks on the Complex Plane. --- Stochastic processes and their Applications 125:2 (2015) 797-818.


\bibitem{Krav2012} A.K. Kravtseva. Feynman integrals of functionals of exponential form with a polynomial exponent. --- Moscow University Mathematics Bulletin 67:5-6 (2012) 233-235


\bibitem{KSS2016} A.K. Kravtseva, O.G. Smolyanov, E.T. Shavgulidze. Asymptotic expansions of Feynman integrals of exponentials with polynomial exponent. --- Russ. J. Math. Phys., 23:4 (2016), 491-509

\bibitem{Maz} S. Mazucchi. Functional-integral solution for the Schr\"odinger equation with polynomial potential: a white noise approach. --- Infin. Dimens. Anal. Quantum. Probab. Relat. Top., 14, 675 (2011). 


\bibitem{KF} A.N. Kolmogorov, S.V.Fomin. Introductory real analysis. --- Dover, 1975.

\bibitem{Naim}M.A. Naimark. Linear differential operators, vol. 2. --- George Harrap Co, London-Toronto-Wellington-Sydney, 1968.

\bibitem{Stone} M.H.Stone. On one-parameter unitary groups in Hilbert Space. --- Annals of Mathematics 33 (3): 643-648, 1932.

\bibitem{EN1} K.-J. Engel, R. Nagel. One-Parameter Semigroups for Linear Evolution Equations. --- Springer, 2000.


\bibitem{R1} I.D. Remizov. Quasi-Feynman formulas -- a method of obtaining the evolution operator for the Schr\"odinger equation. --- J. Funct. Anal. 270:12 (2016), 4540-4557.


\bibitem{Town} John S. Townsend. A Modern Approach to Quantum Mechanics (2-nd Edition). --- University Science Books, 2012.

\bibitem{BerSh} F. A. Berezin, M.A. Shubin. The Schr\"odinger Equation (Series: Mathematics and its Applications, book 66). --- Springer, 1991.

\bibitem{Mull} Harald J. W. Muller-Kirsten. Introduction to Quantum Mechanics: Schr\"odinger Equation and Path Integral (2nd Edition). --- World Scientific, 2012. 

\bibitem{BS} V.I. Bogachev, O.G. Smolyanov. Real and functional analysis: university course. (In Russian) --- M. Izhevsk: RCD, 2009.

\bibitem{Yos} K. Yosida. Functional analysis. --- Springer, 1995.

\bibitem{RS} M. Reed, B. Simon. Methods of Modern Mathematical Physics, vol. 1: Functional Analysis. --- Academic Press, 1980.


\bibitem{RS2} M. Reed, B. Simon. Methods of Modern Mathematical Physics, vol. 2: Fourier Analysis, Self-Adjointness. --- Academic Press, 1975.


\bibitem{HW} A. Hassell and J. Wunsch. The Schr\"odinger Propagator for Scattering Metrics. --- Ann. Math., Vol. 162, No. 1 (Jul., 2005), pp. 487-523.

\bibitem{Nak} S. Nakamura. Wave front set for solutions to Schr\"odinger equations. --- J. Funct. Anal. 256 (2009), pp. 1299–1309.

\bibitem{Ord} G.N.Ord. The Schr\"odinger and diffusion propagators coexisting on a lattice. --- Journal of Physics A: Mathematical and General, Vol. 29, No. 5, 1996. doi:10.1088/0305-4470/29/5/007

\bibitem{IN} K.Ito, S.Nakamura. Remarks on the Fundamental Solution to Schr\"odinger Equation with Variable Coefficients. --- Annales de l’institut Fourier 62.3 (2012): 1091-1121.

\bibitem{WZ}K. Weihrauch, N. Zhong. Is the Linear Schr\"odinger Propagator Turing Computable?. --- Computability and Complexity in Analysis, Lecture Notes in Computer Science, Vol. 2064 (2001), pp. 369-37.

\bibitem{CorN} E. Corderoa, F. Nicola. Some new Strichartz estimates for the Schr\"odinger equation. --- Journal of Differential Equations, Vol. 245, Issue 7 (2008), pp. 1945–1974.


\bibitem{Ah}Y. Aharonov, F. Colombo, I. Sabadini, D.C. Struppa, J. Tollaksen. On the Cauchy problem for the Schr\"odinger equation with superoscillatory initial data. --- Journal de Mathématiques Pures et Appliquées, Vol. 99 (2) Feb. 2013, pp. 165-173.

\bibitem{HMmung} Amru Hussein, Delio Mugnolo. Quantum graphs with mixed dynamics: the transport/diffusion case. --- Journal of Physics A: Mathematical and Theoretical, Volume 46, Number 23, 2013. 

\bibitem{CNR} E. Cordero, F. Nicola and L. Rodino. Gabor representations of evolution operators. --- Trans. Amer. Math. Soc. (2015).

\bibitem{Chernoff} Paul R. Chernoff. Note on product formulas for operator semigroups. --- J. Functional Analysis 2:2 (1968), 238-242. 

\bibitem{STT} O.G. Smolyanov, A.G. Tokarev, A. Truman. Hamiltonian Feynman path integrals via the Chernoff formula. --- J. Math. Phys. 43, 10 (2002) 5161-5171.

\bibitem{SWWdan} Smolyanov O.G., Weiz\"sacker H.v., Wittich O. Diffusion on compact Riemannian manifolds, and surface measures. --- Doklady Math. 2000. V. 61. P. 230-234.

\bibitem{SWWcan} O.G. Smolyanov, H.v. Weizs\"acker, and O. Wittich. Brownian motion on a manifold as limit of stepwise conditioned standard Brownian motions. --- Stochastic processes, Physics and Geometry: New Interplays. II: A Volume in Honour of S. Albeverio, volume 29 of Can. Math. Soc. Conf. Proc., pages 589–602. Am. Math.
Soc., 2000.

\bibitem{STmzm}O. G. Smolyanov, A. Truman. Feynman Formulas for Solutions of the Schrödinger Equation on Compact Riemannian Manifolds. --- Math. Notes, 68:5 (2000), 668-671


\bibitem{SmHist} O.G. Smolyanov. Feynman formulae for evolutionary equations. --- Trends in Stochastic Analysis, London Mathematical Society Lecture Notes Series 353, 2009.

\bibitem{SmSchrHist} O.G.Smolyanov. Schr\"{o}dinger type semigroups via Feynman formulae and all that. Proceedings of the Quantum Bio-Informatics V, Tokyo University of Science, Japan, 7 - 12 March 2011. --- World Scientific, 2013.

\bibitem{F1} R.P. Feynman. Space-time approach to nonrelativistic quantum mechanics. --- Rev. Mod. Phys., 20 (1948), 367-387.

\bibitem{Nel} E. Nelson. Feynman Integrals and the Schr\"odinger Equation. --- J. Math. Phys., 1964, 5(3), p.332.

\bibitem{F2} R.P. Feynman. An operation calculus having applications in quantum electrodynamics. --- Phys. Rev. 84 (1951), 108-128.


\bibitem{AHKM} Sergio Albeverio,Rafael Høegh-Krohn,Sonia Mazzucchi. Mathematical Theory of Feynman Path Integrals: An Introduction. --- Springer, 2008.

\bibitem{Plya1} A.S. Plyashechnik. Feynman formula for Schr\"{o}dinger-Type equations with time- and space-dependent coefficients,
Russian Journal of Mathematical Physics, 2012, vol. 19, No.3, pp. 340-359.

\bibitem{Plya2} A.S. Plyashechnik. Feynman formulas for second-order parabolic equations with variable coefficients. --- 
Russian Journal of Mathematical Physics, 2013, vol. 20, No.3, pp. 377-379.

\bibitem{Butko1} Ya.A. Butko, R.L. Schilling and O.G. Smolyanov. Lagrangian and Hamiltonian Feynman formulae for some Feller semigroups and their perturbations, Inf. Dim. Anal. Quant. Probab. Rel. Top., Vol. 15 N 3 (2012), 26 p.

\bibitem{Butko2} B. B\"{o}ttcher, Ya.A. Butko, R.L. Schilling and O.G. Smolyanov. Feynman formulae and path integrals for some evolutionary semigroups related to tau-quantization, Rus. J. Math. Phys.,  Vol. 18 N 4 (2011), 387-399.


\bibitem{OSS} Yu.N. Orlov, V.Zh. Sakbaev, O.G. Smolyanov. Feynman formulas as a method of averaging random Hamiltonians. --- Proceedings of the Steklov Institute of Mathematics, August 2014, Volume 285, Issue 1, pp 222-232.


\bibitem{R8} I.D.Remizov. Solution-giving formula to Cauchy problem for multidimensional parabolic equation with variable coefficients. --- preprint  	arXiv:1710.06296 (2017)

\bibitem{Dobretal} V.V.Dobrovitski, E.R.Rakhmetov,  B.Barbara, A.K.Zvezdin. Quantum tunnelling of magnetization
in uniaxial magnetic clusters. ---  Acta Phys. Polon. A, 1997, vol. 92, no. 2, pp. 473–476.

\bibitem{OSS-2016} Yu.N. Orlov, V.Zh. Sakbaev, O.G. Smolyanov. Unbounded random operators and Feynman formulae. --- Izvestiya Mathematics, 2016, 80:6, 1131-1158


\bibitem{R2} I.D. Remizov. Solution of a Cauchy problem for a diffusion equation in a Hilbert space by a Feynman formula. --- Russian Journal of Mathematical Physics, 2012, v.19, No.3, 360-372.

\bibitem{SWW}O.G. Smolyanov, H. v. Weizs\"{a}cker, O. Wittich. Chernoff's Theorem and Discrete Time Approximations of Brownian Motion on Manifolds. --- Potential Analysis, February 2007, Volume 26, Issue 1, pp 1-29.


\bibitem{BGS2010} Ya.A. Butko, M. Grothaus, O.G. Smolyanov. Lagrangian Feynman formulas for second-order parabolic equations in bounded and unbounded domains. --- Infinite Dimansional Analyasis, Quantum Probability and Related Topics, vol. 13, No. 3 (2010), 377-392.


\bibitem{R3} I.D. Remizov.  Solution of the Schr\"odinger Equation with the Use of the Translation Operator. --- Math. Notes, 100:3 (2016), 499-503

\bibitem{Dubravina} V.\,A.~Dubravina. Feynman formulas for solutions of evolution equations on ramified surfaces. --- Russian Journal of Mathematical Physics
Apr. 2014, Vol. 21 (2) pp. 285-288.

\bibitem{Bmnog} Ya.A.Butko. Feynman formulas and functional integrals for diffusion with drift in a domain on a manifold. --- Mathematical Notes, April 2008, Volume 83, Issue 3-4, pp 301-316

\bibitem{R4} I.D. Remizov. On the connection between the resolving semigroups and the families of operators Chernoff-equivalent to them for the heat and the Schr\"{o}dinger equations in $L^2$ space (in Russian). --- Proceedings of the Lomonosov-2014 conference, Moscow State University, April 2014.

\bibitem{SYaa} V. Sakbaev and A. Yaakbarieh. Feynman Formulas Representation of Semigroups Generated by Parabolic Difference-Differential Equations. --- American Journal of Computational Mathematics, Vol. 2 No. 4, 2012, pp. 295-301.

\bibitem{SShav} O.G. Smolyanov, E.T. Shavgulidze. Continual integrals (in Russian). --- URSS, 2015, ISBN 978-5-9710-2133-9

\bibitem{R6} I.D. Remizov. Quasi-Feynman formulas provide solution for multidimensional Schr\"odinger equation with unbounded potential. --- Mathematical Notes, accepted for publication, to be published in  2018.

\bibitem{SakTMF} V. Zh. Sakbaev. Averaging of random walks and shift-invariant measures on a Hilbert space. --- Theoret. and Math. Phys., 191:3 (2017), 886-909.

\bibitem{Buz2017} M.S. Buzinov. Feynman Formulas for Semigroups
Generated by an Iterated Laplace Operator. --- Russian Journal of Mathematical Physics, Vol. 24, No. 2, 2017, pp. 250-255

\bibitem{Buz2015} M.S. Buzinov. Feynman and Quasi-Feynman formulae for evolution equations with a polyharmonic Hamiltonian. --- Int. Conf. "Infinite-dimensional dynamics, dissipative systems, and attractors", Nizhny Novgorod (Russia), July 13-17, 2015.


\bibitem{R5} I.D. Remizov. New Method for Constructing Chernoff Functions. --- Differential Equations, 53:4 (2017), 566-570

\bibitem{SSWW} Nadezda A. Sidorova, Oleg G. Smolyanov, Heinrich v. Weizs\"acker, Olaf Wittich. 
The surface limit of Brownian motion in tubular neighborhoods of an embedded Riemannian manifold. --- Journal of Functional Analysis, 206:2, 253-500, 2004.

\bibitem{R7} I.D.Remizov. Feynman  and Quasi-Feynman formulas for Evolution Equations. --- Doklady Mathematics 96:2 (2017) pp. 433-437

\bibitem{R-AMC} Ivan D. Remizov. Approximations to the solution of Cauchy problem for a linear evolution equation via the space shift operator (second-order equation example). --- Applied Mathematics and Computaton 328 (2018), 243-246.

\bibitem{Zag-1} V.A.Zagrebnov. Comments on the Chernoff $\sqrt{n}$-lemma. In book [Series of Congress Reports: Functional Analysis and Operator Theory for Quantum Physics, The Pavel Exner Anniversary Volume. Jaroslav Dittrich, Hynek Kova\v{r}\'ik, Ari Laptev (eds). European Mathematical Society, 2017] pp. 564-573, 2017.

\bibitem{Zag-2} Hagen Neidhardt,
Artur Stephan and Valentin A. Zagrebnov. Remarks on the operator-norm convergence of the Trotter product
formula. --- Integral Equations and Operator Theory, 90:15 (2018). 

\bibitem{Zag-3} Hagen Neidhardt, Artur Stephan, Valentin A. Zagrebnov. Operator-Norm Convergence of the Trotter Product Formula on Hilbert and Banach Spaces: A Short Survey. --- Current Research in Nonlinear Analysis, pp 229-247 (2018).




\end{thebibliography}
\end{document}